\documentclass[showpacs,aps,prstab,amsmath,amssymb]{revtex4-1}
\usepackage[latin9]{inputenc}
\usepackage{verbatim}
\usepackage{textcomp}
\usepackage{amsmath}
\usepackage{graphicx}
\usepackage{esint}

\makeatletter

\DeclareFontEncoding{LGR}{}{}
\DeclareRobustCommand{\greektext}{%
  \fontencoding{LGR}\selectfont\def\encodingdefault{LGR}}
\DeclareRobustCommand{\textgreek}[1]{\leavevmode{\greektext #1}}
\ProvideTextCommand{\~}{LGR}[1]{\char126#1}

\InputIfFileExists{t2aenc.def}{}{%
  \errmessage{File `t2aenc.def' not found: Cyrillic script not supported}}

\newcommand{\lyxmathsym}[1]{\ifmmode\begingroup\def\b@ld{bold}
  \text{\ifx\math@version\b@ld\bfseries\fi#1}\endgroup\else#1\fi}

\DeclareTextSymbolDefault{\textquotedbl}{T1}
\newcommand{\lyxdot}{.}

\usepackage{bm}
\usepackage{booktabs}
\usepackage[usenames]{color}
\usepackage{hyphenat}

\renewcommand{\[}{\begin{equation}}
\renewcommand{\]}{\end{equation}}
\makeatother

\begin{document}
\title{A Concept of Two-Point Propagation Field of a Single Photon: A Way
to X-ray Picometer Displacement Detection and Nanometer Resolution
3D X-ray Micro-Tomography}
\author{Li Hua Yu}
\affiliation{NSLSII, Brookhaven National Laboratory}
\begin{abstract}
We introduce the two-point propagation field (TPPF) \textemdash{}
a real-valued, phase-sensitive quantity defined as the functional
derivative of the single-photon detection probability with respect
to an infinitesimal opaque perturbation placed between source and
detection slits. The TPPF is analytically derived and shown to exhibit
stable, high-frequency sinusoidal structure with periods reaching
4\textendash 7 nm near the detection slit for X-rays. This structure
enables shot-noise-limited displacement detection at $\sim200$ pm
precision for 6 keV X-rays using total photon counts on the order
of $1\times10^{7}$ and detector photon counting as low as 287. Beyond
displacement detection, the TPPF physically performs a Fourier-Radon
transformation of the projection data, providing a pathway to non-iterative
frequency-domain tomography. Two conceptual strategies\textbackslash textemdash
a central blocker and off-axis multi-slit arrays\textbackslash textemdash
are estimated to lower the required incident photon budget by more
than one order of magnitude each, yielding combined reductions of
two to three orders of magnitude with near-term detector development.
The TPPF concept, originally developed in a perturbative study of
single-particle propagation, bridges quantum measurement questions
with practical high-resolution X-ray physics. This work provides the
foundational physics required for future discrete sampling and 3D
numerical reconstruction algorithms.
\end{abstract}
\maketitle
keyword: X-ray wave optics; Phase tomography; Radon transform; Inverse
problems;Fourier optics; X-ray interferometry;Picometer displacement
detection

\section*{1. Introduction and motivation}

We outline a concept\textemdash the two-point propagation field (TPPF)\textemdash that
may provide a practical basis for achieving shot-noise-limited X-ray
displacement detection with \ensuremath{\sim} 200 pm precisions for
existing synchrotron beamlines and practical slit-comb samples. This
approach addresses the relative sample-beam motion that restricts
resolution to \ensuremath{\sim} 4 nm \cite{swiss_tomography_4nm}.
The method is based on the two-point propagation field (TPPF), a real-valued,
phase-sensitive function derived from perturbative analysis of single-particle
propagation. TPPF is derived from the perturbative response of detection
rates to localized perturbations\textemdash such as placing a thin
opaque pin to block the wave\textemdash at intermediate points along
the photon propagation path between a source slit and a detection
slit. This quantity, referred to as the two-point propagation field
(TPPF), was introduced and derived in detail in a recent theoretical
study \cite{yu-1}, where it was originally termed the \textquotedbl perturbative
function''.

The central idea is that this function captures projection-like information
about a sample\textquoteright s internal structure by encoding how
the count rate at a detector changes when the wave is perturbed by
a pin at different points. When we replace the perturbation pin with
a sample as the perturbation, the detection rate becomes a function
of the position and the angle of the sample orientation. The function
is a convolution between the TPPF and the sample's structural function.
The result is a stable, reproducible function that varies with the
scan position and angle of the sample, and may be viewed as analogous
to the projection data used in Radon-transform-based reconstruction
techniques. However, in this case, the measured signal is directly
related to the Fourier transform of the Radon transform, due to the
high-frequency phase information inherent in the TPPF.

Although this work does not address reconstruction techniques, the
structure of the TPPF suggests a possible path toward extending classical
tomographic methods to single-particle quantum systems\textemdash particularly
in efforts to resolve nanometer-scale structural features. This note
presents the TPPF formula, its physical interpretation, and a proposed
connection to projection-based 3D-tomography \cite{3Dtomo_flann,3Dtomo_path}.
The objective of this work is to establish a clear framework for researchers in 3D imaging and X-ray science, and to explore whether the method can be experimentally tested and integrated into an existing tomographic method. As the first step in this direction, the analysis leads to
practical shot-noise-limited displacement detection precision of \ensuremath{\sim}200
pm for existing synchrotron beamlines and practical comb/slit geometries.
Such precision enables lensless, counting-based detection that requires
stability only over the final 0.5 mm propagation distance, utilizing
total photon counts of $10^{6}$ to $10^{7}$, readily achievable
at synchrotron or XFEL beamlines, depending on slit configurations.

This step itself is useful for the advancement of X-ray tomography
because the motion between the X-ray beam and the sample of order
of 4 nm is one of the limitations of tomography resolution\cite{swiss_tomography_4nm}.
By encoding high-frequency phase information akin to a Fourier-transformed
Radon projection (Section 3), the TPPF supports reconstruction of
internal structures at nanometer scales, offering a pathway to lower-dose
imaging compared to conventional ptychography or burst methods. This
not only facilitates experimental validation of the TPPF but also
reduces radiation damage in biological samples, potentially by more
than one order of magnitude, through strategies like central blockers
or off-axis slit arrays, as explored in Section 5.

Section 2 introduces the two-point propagation function (TPPF). Section
3 examines its connection to the Radon transform and 3D tomography.
In Section 4, we illustrate how to apply the calculation steps for
the X-ray picometer displacement detection to an example in a 6 keV
X-ray case. Section 5 describes ongoing work to significantly reduce
the photon count requirements for achieving the target sensitivity.
Section 6 discusses the relation between TPPF and the quantum measurement
during the free-space propagation between the source and the detector
slit. Section 7 is the conclusion.

This work is approached from a theoretical standpoint, using broadly
referenced parameters to assess the compatibility of a picometer X-ray
displacement detection and nanometer-resolution 3D tomography with
existing technology. Although not based on direct experience with
experimental tomography or biological imaging, the analysis intends
to provide a foundation for shared exploration and further evaluation
by experts in these areas.

\begin{comment}
Fourier transform of the Radon-transform: see/Users/lihuayu/Desktop/QMwave\_collapse/no\_bath\_free\_space/references/tomography\_reconstruction/ref.lyx/ref1
\end{comment}

\section*{2. Description of the experiment to measure the two-point propagation
field}

\subsection*{2.1 TPPF and its physical interpretation}

\begin{figure*}
\includegraphics[viewport=2bp 2bp 698bp 449bp,width=0.55\columnwidth]{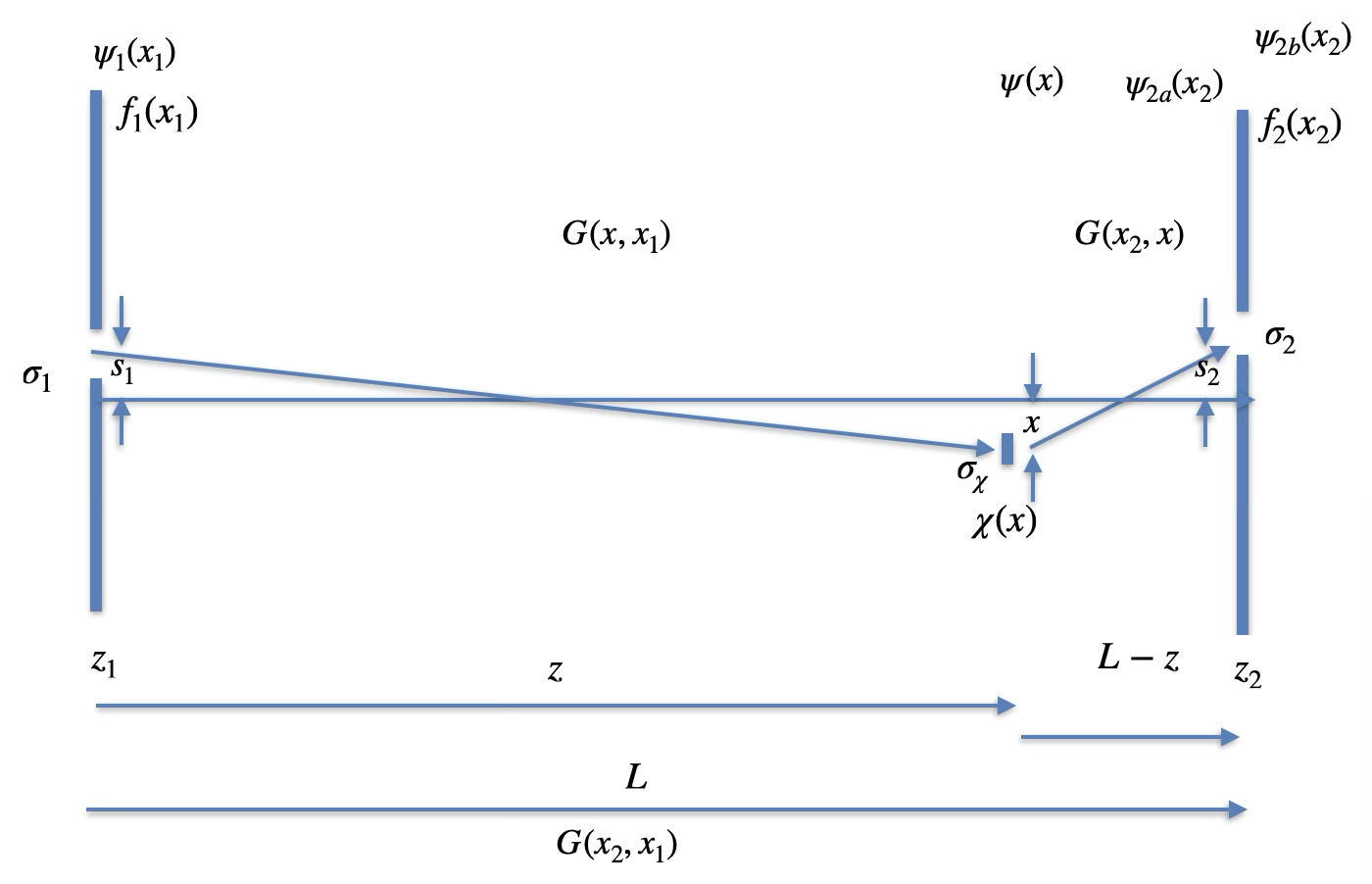}\vspace{1.0em}\includegraphics[viewport=0bp 0bp 220bp 258bp,width=0.3\columnwidth]{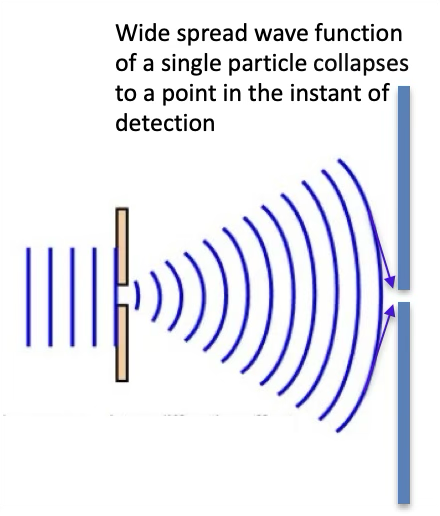}\vspace{-1.0em}

\caption{(a) Experimental geometry $\textbf{(not to scale)}$. The wave functions
at slits 1,2 at $z_{1},z_{2}$ with apertures $\sigma_{1},\sigma_{2}$
and corresponding transverse displacements $s_{1},s_{2}$ in the x-direction,
and a pin $\chi(x)$ at $z,x_{p}$ of width $\sigma_{\chi}$. The
distances between the slits and the pin are $z,L-z,$ and $L.$ The
longitudinal axis is the $z$-axis, the x-axis is vertical in this
figure. The slits are perpendicular to the plane of the figure, parallel
to the $y$-axis. We use $\psi_{1}(x_{1},z_{1})$, $\psi(x,z),$$\psi_{2a}(x_{2},z_{2})$
and $\psi_{2b}(x_{2},z_{2})$ to represent the wave function at the
entrance, the pin, and the exit, respectively. The subscripts $a$
and $b$ represent before and after detector slit 2 . $f_{1}(x_{1})$
and $f_{2}(x_{2})$ represent the effect of the slits such that $\psi_{1}(x_{1},z_{1})\equiv f_{1}(x_{1}),\psi_{2b}(x_{2},z_{2})=f_{2}(x_{2})\psi_{2a}(x_{2},z_{2})$.
If we choose the slit with the hard-edged opening, $f_{1}(x_{1})$
and $f_{2}(x_{2})$ would be zero outside the slits and equal to 1
within the slits. To simplify the calculation, we assume they are
Gaussian with peak value 1, except that we choose $f_{1}(x_{1})$
to normalize $\psi_{1}$ as $P_{1}=\int dx_{1}|\psi_{1}(x_{1},z_{1}=0)|^{2}=1$.
The pin profile is $\chi(x)=1$ when it is removed. When inserted,
$\chi(x)=1-\exp(-\frac{1}{2\sigma_{\chi}^{2}}\left(x-x_{p}\right)^{2})$;
the effective width (equivalent hard-edged slit width) is $\Delta x=\sqrt{2\pi}\sigma_{\chi}$.\label{fig:(a)-setup}
(b)The wave function of a single particle spreads over a wide region
after emission and collapses instantaneously upon detection\label{fig:(b)-The-pond-wave}}
\vspace{-1.0em}
\end{figure*}

As illustrated in Figure \ref{fig:(a)-setup}(a), we analyze the wave
function evolution of a particle when it propagates through free space
in the longitudinal $z$-direction from a thin entrance slit 1 to
a detector behind a thin exit detector slit 2  to find the information
about whether the wave function collapse occurs at the entrance slit
1 or the exit detector slit 2 . The slits are parallel to the $y$-axis
(perpendicular to the plane of the figure). The $x$-axis is vertical
in the figure. Between the slits, the wave function must follow the
Schr{\"{o}}dinger equation because the only non-unitary (irreversible) process
is at the slits. Only the particles that pass through the slits are
selected and detected. The probability of a particle found between
$z_{1}$ and $z_{2}$ is a constant independent of $z$ due to particle
number conservation. We insert a thin pin $\chi(x)=1+\Delta\chi(x)$
between the two slits at position $z,x$ to cut off the wave function
as a perturbation. When $\Delta\chi(x)=0$, there is no pin. When
$\Delta\chi(x,z)=-1$ within a thin section $\Delta x$ around a point
at (x,z), it represents a thin pin. When $|\Delta\chi(x,z)|\ll1$
in an extended area, it becomes a sample as a perturbation. The caption
for Fig. 1(a) gives the notations.

In Fig.\ref{fig:(a)-setup}(a), we study the effect of a perturbation
$\Delta\chi(x)$ on the counting rate $P_{2b}=\int_{-\infty}^{\infty}dx_{2}|\psi_{2b}(x_{2})|^{2}$,
and calculate the ratio of the counting rate change over the perturbation.
As the perturbation approaches zero, the ratio $\frac{\Delta P_{2b}}{\Delta\chi(x,z)}$
becomes the functional derivative $\frac{\delta P_{2b}}{\delta\chi(x,z)}$
of the counting rate over the perturbation $\Delta\chi(x)$. This
perturbative function is independent of perturbation; it is a real-valued
function containing high-resolution phase information, determined
solely by the two-slit geometry in our 2D study, as demonstrated later.
It can be measured with high precision and reproducibility, and it
uniquely characterizes the individual event of a particle propagating
between the two slits. Unlike a probability amplitude, it manifests
as a stable, reproducible structure corresponding to a physically
meaningful propagation quantity exhibiting high-resolution fringes,
which we define as the two-point propagation field (TPPF). The TPPF
does not correspond to a probability distribution. While the wave
function describes an ensemble of possible detection outcomes, the
TPPF characterizes the process underlying a single detection \textemdash{}
a realization selected according to the Born rule.

In Appendix II we outline a derivation of the calculation of TPPF
using the basics of quantum mechanics. Since in this work we consider
X-ray 3D tomography, for the case in Fig.\ref{fig:(a)-setup} (a)
and for sufficiently long and narrow slits and the pin, we neglect
the $y$ dependence and assume a 2D Maxwell equation

\begin{equation}
\frac{1}{c^{2}}\frac{\partial^{2}}{\partial t^{2}}\phi(x,z,t)=\left(\frac{\partial^{2}}{\partial x^{2}}+\frac{\partial^{2}}{\partial z^{2}}\right)\phi(x,z,t).\label{eq:1}
\end{equation}

Because we do not consider polarization, we just use $\phi$ to denote
any component of the electromagnetic field. We assume $k=\frac{\omega}{c}$,
$\phi(x,z,t)=e^{i(kz-\omega t)}\psi(x,z)$, take the paraxial approximation:
\textbar$\frac{\partial^{2}}{\partial z^{2}}\psi|\ll|\frac{\partial^{2}}{\partial x^{2}}\psi$\textbar ,
\textbar$\frac{\partial^{2}}{\partial z^{2}}\psi|\ll2k|\frac{\partial}{\partial z}\psi$\textbar ,
we get the 1D Schr{\"{o}}dinger equation $i\frac{\partial}{\partial z}\psi=-\frac{1}{2k}\frac{\partial^{2}}{\partial x^{2}}\psi$.
$\psi$ is the slowly varying amplitude and phase of the wave function.
We can characterize the experiment by specifying the wavelength $\lambda=\frac{2\pi}{k}$
only. The Green's function from entrance slit 1 to detector slit 2
 is

\begin{align}
 & G(x_{2},x_{1};z_{2},z_{1})=\left(\frac{k}{2\pi i(z_{2}-z_{1})}\right)^{\frac{1}{2}}\exp\left[i\frac{k}{2(z_{2}-z_{1})}\left(x_{2}-x_{1}\right)^{2}\right]\label{greens function}
\end{align}

The initial wave function $\psi_{1}(x_{1},z_{1})\equiv f_{1}(x_{1})=\left(\frac{1}{2\pi\sigma_{1}^{2}}\right)^{\frac{1}{4}}\exp(-\frac{1}{4\sigma_{1}^{2}}(x_{1}-s_{1})^{2})$
is the Gaussian profile of entrance slit 1, $f_{2}=\exp(-\frac{1}{4\sigma_{2}^{2}}(x_{2}-s_{2})^{2})$
is for detector slit 2 . Our analysis shows that for the narrow slits
we considered, replacing the Gaussian profile by a hard-edge profile
of width $\sqrt{2\pi}\sigma_{2}$ would only cause a negligible difference.
The Gaussian profile allows us to derive a simple analytical result,
as described in Section 2.2.

In the derivation of the TPPF, the Green's functions $G(x,x_{1}$),$G(x_{2},x$)
are given by replacing the corresponding variables in Eq.(\ref{greens function}).
When wave function at the entrance slit 1 is normalized as $\psi_{1}(x_{1},z_{1})\equiv f_{1}(x_{1})$,
applying the property of the Green's function, without the pin, the
wave function at the exit of detector slit 2  is $\psi_{2b}(x_{2})=f_{2}(x_{2})\int_{-\infty}^{\infty}dxG(x_{2},x;t_{2}-t)\int_{-\infty}^{\infty}dx_{1}G(x,x_{1};t-t_{1})f_{1}(x_{1})$.
When there is a perturbation $\chi(x)=1+\Delta\chi(x)$ of a pin,
the wave function after the detector slit 2  is

\begin{equation}
\psi_{2b}(x_{2})+\Delta\psi_{2b}(x_{2})=f_{2}(x_{2})\int_{-\infty}^{\infty}dxG(x_{2},x;t_{2}-t)\left(1+\Delta\chi(x)\right)\int_{-\infty}^{\infty}dx_{1}G(x,x_{1};t-t_{1})f_{1}(x_{1})\label{pertubation:}
\end{equation}

Because $\Delta\chi(x)=-1$ within the thin section $\Delta x,$ the
increment for infinitesimal $\Delta x$ is 

\begin{align*}
 & \Delta\psi_{2b}(x_{2})=f_{2}(x_{2})\int_{-\infty}^{\infty}dxG(x_{2},x;t_{2}-t)\Delta\chi(x)\int_{-\infty}^{\infty}dx_{1}G(x,x_{1};t-t_{1})f_{1}(x_{1})\\
 & =-\int_{x}^{x+\Delta x}dxf_{2}(x_{2})G(x_{2},x;t_{2}-t)\int_{-\infty}^{\infty}dx_{1}G(x,x_{1};t-t_{1})f_{1}(x_{1})\\
 & =-\Delta xf_{2}(x_{2})G(x_{2},x;t_{2}-t)\int_{-\infty}^{\infty}dx_{1}G(x,x_{1};t-t_{1})f_{1}(x_{1})
\end{align*}

Thus, the functional derivative of $\psi_{2b}(x_{2})$ over the perturbation
is

\begin{align}
 & \frac{\delta\psi_{2b}(x_{2})}{\delta\chi(x,z)}=\frac{\Delta\psi_{2b}(x_{2})}{\Delta x}=-f_{2}(x_{2})G(x_{2},x;t_{2}-t)\int_{-\infty}^{\infty}dx_{1}G(x,x_{1};t-t_{1})f_{1}(x_{1})\label{psi_derivative}
\end{align}

The TPPF is the functional derivative of the detection probability
$P_{2b}=\int_{-\infty}^{\infty}dx_{2}|\psi_{2b}(x_{2})|^{2}$ over
the perturbation $\Delta\chi(x)$, and we define the \textbf{complex
part} of TPPF as $\frac{\delta P_{2b}^{(c)}}{\delta\chi(x,z)}\equiv\int_{-\infty}^{\infty}dx_{2}\frac{\delta\psi_{2b}(x_{2})}{\delta\chi(x,z)}\psi_{2b}^{*}(x_{2})$,
we have,

\begin{align}
 & \frac{\delta P_{2b}}{\delta\chi(x,z)}=\int_{-\infty}^{\infty}dx_{2}\frac{\delta\psi_{2b}(x_{2})}{\delta\chi(x,z)}\psi_{2b}^{*}(x_{2})+\int_{-\infty}^{\infty}dx_{2}\psi_{2b}(x_{2})\frac{\delta\psi_{2b}^{*}(x_{2})}{\delta\chi(x,z)}\equiv\frac{\delta P_{2b}^{(c)}}{\delta\chi(x,z)}+c.c.\label{dP2bdchi-1}
\end{align}

Applying Eq.(\ref{psi_derivative}) , we have the complex TPPF

\begin{align}
 & \frac{\delta P_{2b}^{(c)}}{\delta\chi(x,z)}=-\int_{-\infty}^{\infty}dx_{2}\int_{-\infty}^{\infty}dx_{1}\psi_{2b}^{*}(x_{2})f_{2}(x_{2})G(x_{2},x;t_{2}-t)G(x,x_{1};t-t_{1})f_{1}(x_{1})\nonumber \\
 & =\int_{-\infty}^{\infty}dx_{2}f_{2}^{2}(x_{2})G(x_{2},x;t_{2}-t)\int_{-\infty}^{\infty}dx_{1}G(x-x_{1},x_{1};t-t_{1})f_{1}(x_{1})\int_{-\infty}^{\infty}dx'_{1}G^{*}(x_{2}-x'_{1};t_{2}-t_{1})f_{1}(x'_{1})\label{complex_dpdchi}
\end{align}

The last step shows there is a simple relation between TPPF and the
probability $P_{2b}$, i.e. 

\begin{align}
 & \int_{-\infty}^{\infty}dx\frac{\delta P_{2b}^{(c)}}{\delta\chi(x,z)}=\int_{-\infty}^{\infty}dx_{2}f_{2}^{2}(x_{2})\left(\int_{-\infty}^{\infty}dxG(x_{2},x;t_{2}-t)\int_{-\infty}^{\infty}dx_{1}G(x-x_{1},x_{1};t-t_{1})f_{1}(x_{1})\right)\psi_{2a}^{*}(x_{2})\nonumber \\
 & =\int_{-\infty}^{\infty}dx_{2}f_{2}^{2}(x_{2})\psi_{2a}(x_{2})\psi_{2a}^{*}(x_{2})=\int_{-\infty}^{\infty}dx_{2}\psi_{2b}(x_{2})\psi_{2b}^{*}(x_{2})=P_{2b}\nonumber \\
 & P_{2b}=\frac{1}{2}\int_{-\infty}^{\infty}dx\frac{\delta P_{2b}^{(c)}}{\delta\chi(x,z)}+c.c.=\frac{1}{2}\int_{-\infty}^{\infty}dx\frac{\delta P_{2b}}{\delta\chi(x,z)}\label{P2b_TPPF_relation}
\end{align}

Thus, the integration of TPPF over $x$ is not the probability but
twice of it. An important observation is that $\frac{1}{2P_{2b}}\frac{\delta P_{2b}}{\delta\chi(x,z)}$
is a function of $z$, i.e. the position $z$ of the pin, but $\frac{1}{2P_{2b}}\int_{-\infty}^{\infty}dx\frac{\delta P_{2b}}{\delta\chi(x,z)}=1$
is independent of $z$ and represents the particle number conservation
during the free propagation in free space. Although the two-point
propagation field (TPPF) is not positive-definite, its integral over
space yields a conserved total probability. When normalized and scaled
by $h\nu$, it acquires the dimensions of energy density and integrates
to the particle\textquoteright s total energy. The possibility of
local negative values is not without precedent; similar behavior occurs
in quantum field theory, such as in the Casimir effect \cite{casimir}.
We therefore interpret the scaled TPPF as a generalized energy density
associated with the spatial structure of single-particle propagation.

Since $\frac{1}{2P_{2b}}\frac{\delta P_{2b}}{\delta\chi(x,z)}$ is
the real part of the complex-valued $\frac{\delta P_{2b}^{(c)}}{\delta\chi(x,z)}$,
its imaginary part\textemdash and thus the full complex derivative\textemdash can
be readily obtained via a Hilbert transform\cite{hilbert_tranform}
by multiplying the coefficients of the Fourier transform of $\frac{1}{2P_{2b}}\frac{\delta P_{2b}}{\delta\chi(x,z)}$
by a constant according to the sign of the frequency of the term (see
Appendix I) and followed by an inverse Fourier transform. As we shall
show, $\psi(x,z)$ and $\frac{1}{2P_{2b}}\frac{\delta P_{2b}}{\delta\chi(x,z)}$
are entirely different functions: $\psi(x,z)$ is wide spread as illustrated
in Fig.\ref{fig:(b)-The-pond-wave}(b), however $\frac{1}{2P_{2b}}\frac{\delta P_{2b}}{\delta\chi(x,z)}$
becomes widespread only between the two slits and finally converges
into the detector slit 2 . This answered the fundamental question:
the evolution of the energy distribution $h\nu$ is continuous without
sudden collapse at the detector, as suggested by the TPPF model, even
though the wave function $\psi(x)$ collapses at the instant of the
detection. $\psi(x)$ represents a statistical distribution of the
ensemble, while TPPF$\frac{1}{2P_{2b}}\frac{\delta P_{2b}}{\delta\chi(x,z)}$
represents an individual realization of the ensemble.

\subsection*{2.2 The expression of TPPF and its difference from the wave function
$\psi(x)$}

This simple relation between TPPF and counting rate in Eq.(\ref{P2b_TPPF_relation})
is used in simplifying the derivation of the complex TPPF Eq.($\ref{complex_dpdchi}$)
in Appendix II. The result for the counting rate (i.e., the probability)
$P_{2b}$ and the explicit complex valued TPPF $\frac{\delta P_{2b}^{(c)}}{\delta\chi(x,z)}$
is simpler when expressed in terms of a few scaled parameters in Fig.(\ref{fig:(a)-setup})
for slits width, and longitudinal position: $\mu\equiv\frac{4\pi\sigma_{1}^{2}}{\lambda L}=\frac{2k\sigma_{1}^{2}}{L}$,
$\rho\equiv\frac{\sigma_{2}^{2}}{\sigma_{1}^{2}},$ $z_{2}=L,$ $z_{1}=0,$
$z=\xi L$:

\begin{align}
 & P_{2b}=\sqrt{\frac{\mu^{2}\rho}{\mu^{2}+\rho\mu^{2}+1}}\exp\left(-\frac{1}{2\sigma_{1}^{2}}{\displaystyle \frac{\mu^{2}}{\mu^{2}+1+\mu^{2}\rho}}(s_{1}-s_{2})^{2}\right)\nonumber \\
 & \frac{1}{P_{2b}}\frac{\delta P_{2b}^{(c)}}{\delta\chi(x,z)}=\frac{\sqrt{-\alpha_{\chi}}}{\sqrt{\pi}}\exp\left({\displaystyle \alpha_{\chi}\left(x-x_{c}\right)^{2}}\right)\label{TPPF}\\
 & \alpha_{\chi}=-\frac{1}{2\sigma_{1}^{2}}\frac{i\mu{\displaystyle \left(\mu^{2}+\rho\mu^{2}+1\right)}}{\left(-i\mu+\xi\right)\left(\mu{\displaystyle \left(i\xi-\mu\right)}\rho+2\left(\xi-1\right)\left(i\mu+1\right)\right)}\nonumber 
\end{align}

Here $x_{c}=\frac{{\displaystyle c_{S1}}s_{1}+{\displaystyle c_{S2}}s_{2}}{{\displaystyle \mu^{2}+\rho\mu^{2}+1}},$
$s_{1},s_{2}$ are the transverse displacement of the slits in Fig.\ref{fig:(b)-The-pond-wave}(a),
and ${\displaystyle c_{S1}}\equiv{\displaystyle \rho\mu^{2}-\left(i\mu+1\right)}\left(\xi-1\right)$,
${\displaystyle c_{S2}}\equiv{\displaystyle \left(\mu-i\right)\left(\mu+i\xi\right)}$.
For the next discussion on the tomography, we only consider the case
of $s_{1}=s_{2}=0$, thus $x_{c}=0$, and hence TPPF is characterized
mainly by the probability $P_{2b}$ and $\alpha_{\chi}$ that provide
the information about the distribution width, fringe spacing, and
frequency bandwidth. etc. of TPPF.

\begin{figure*}
\includegraphics[width=0.4\columnwidth]{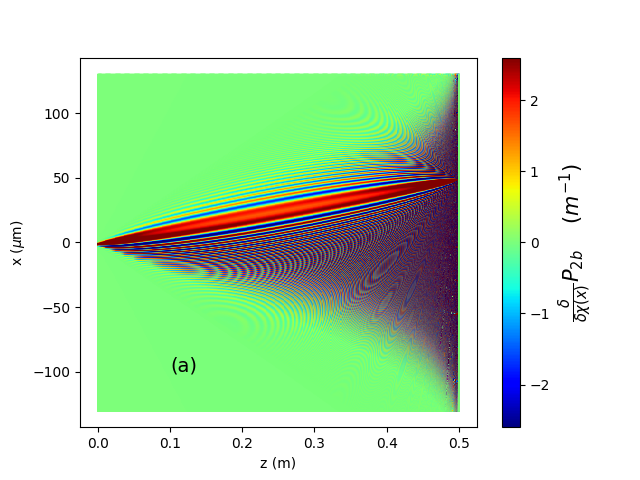}\vspace{0.0em}\includegraphics[width=0.4\columnwidth]{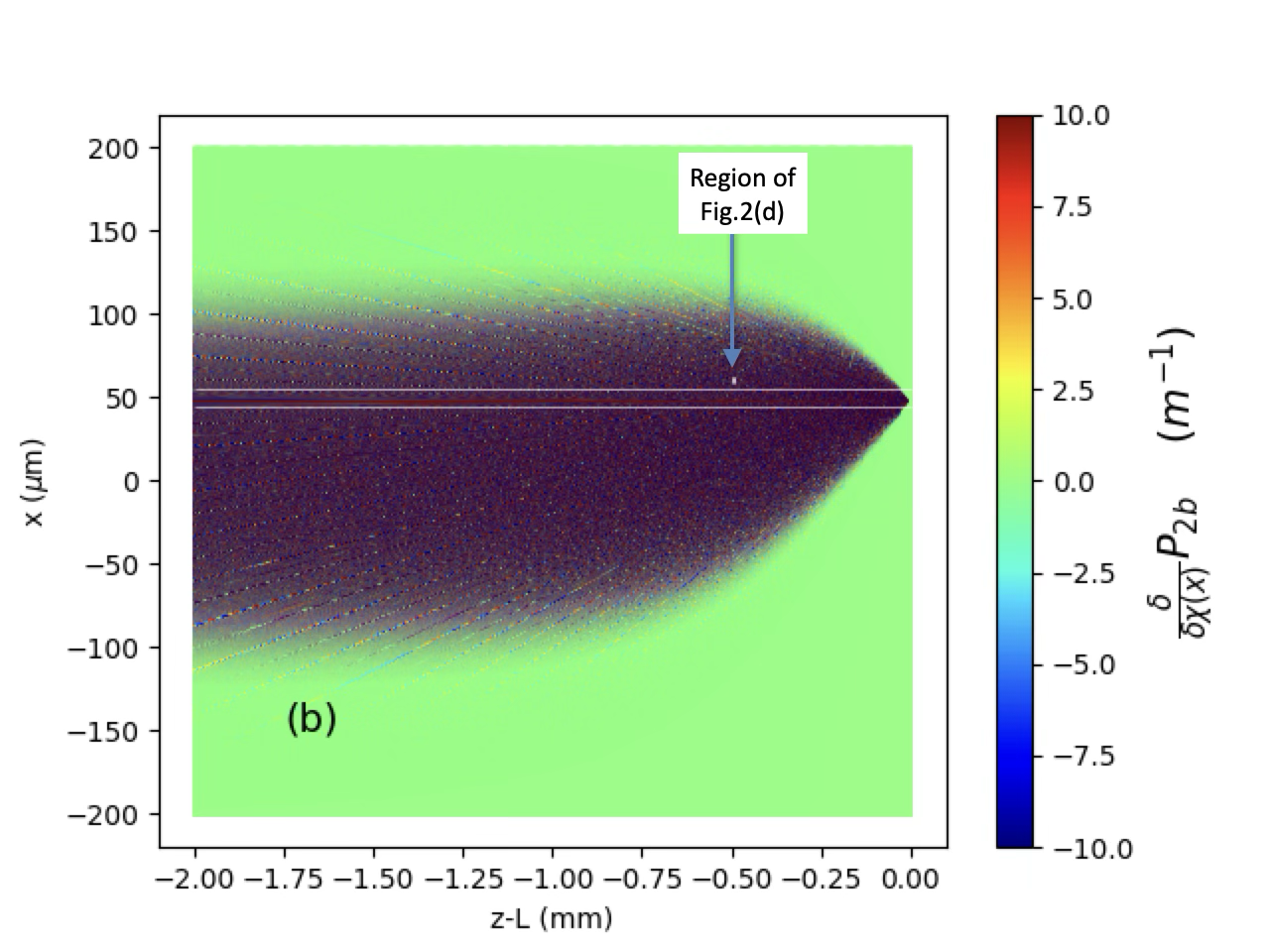}\vspace{0.0em}

\includegraphics[width=0.4\columnwidth]{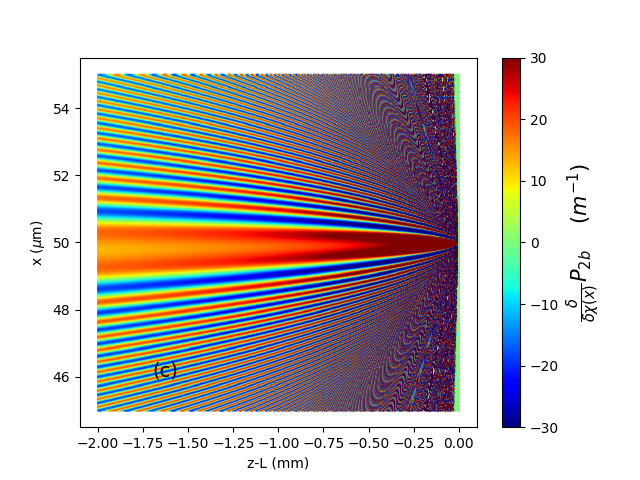}\vspace{0.0em}\includegraphics[width=0.4\columnwidth]{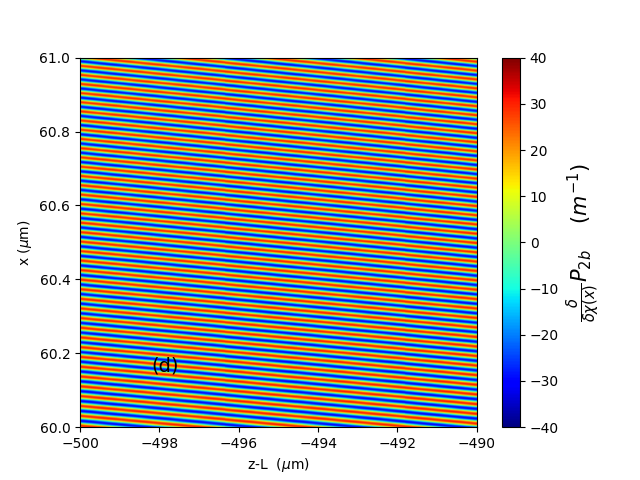}\vspace{0.0em}

\caption{For a setup in Fig.\ref{fig:(a)-setup}(a), take $\lambda=0.541\text{nm},\sigma_{2}=0.8\text{nm},\sigma_{1}=0.5\mu\text{m},L=0.5m,L-z=0.5\text{mm}$,
Fig.2(a): $\frac{\delta P_{2b}}{\delta\chi(x,z)}$ vs. $x,z$ in color
scale for $s_{2}=50\text{\ensuremath{\mu}}m$, $P_{2b}=9.47\times10^{-6}$.
Some elliptical patterns are artifacts due to the limited number of
points of the plot and the nearly periodic structure of the function
$\frac{\delta P_{2b}}{\delta\chi(x,z)}$. The patterns change with
the number of points of the plot, but it is hard to avoid, even with
pixels increased to $4\times10^{6}$ in the plot. Fig.2(b): A narrow
region within 2 mm from the detector slit 2 in (a) showing the details
not visible in (a). The detailed fringe structure is not visible in
this plot because it is visible only when magnified, as given in the
following (c) and (d) plots. The region between the two white lines
(45$\mu\text{m}<x<55\text{\ensuremath{\mu\text{m}}})$ is given in
(c) with details. The hardly visible white dot, which is too small
to be recognized as a box, indicates the region (pointed to by the
arrow in Fig.2(b)) shown in (d) with fringe details. Fig.2(c): The
region (45$\mu\text{m}<x<55\text{\ensuremath{\mu\text{m}}})$ indicated
by the two white lines in (b). Fig.2(d): The region indicated in (b)
by an arrow as a white dot in a box size of $10\mu\text{m}\times1\mu\text{m}$
(60$\mu\text{m}<x<61\text{\ensuremath{\mu\text{m},-500<z-L<-490\text{\ensuremath{\mu\text{m}}})}}$
shows the fringe structure. The most pronounced feature is that the
amplitude ($\pm30\text{m}^{-1})$ indicated by the color scale is
comparable to the peak amplitude in Fig.2(c). \label{fig:dPdx_color}}
\end{figure*}

\begin{figure*}
\includegraphics[width=0.4\columnwidth]{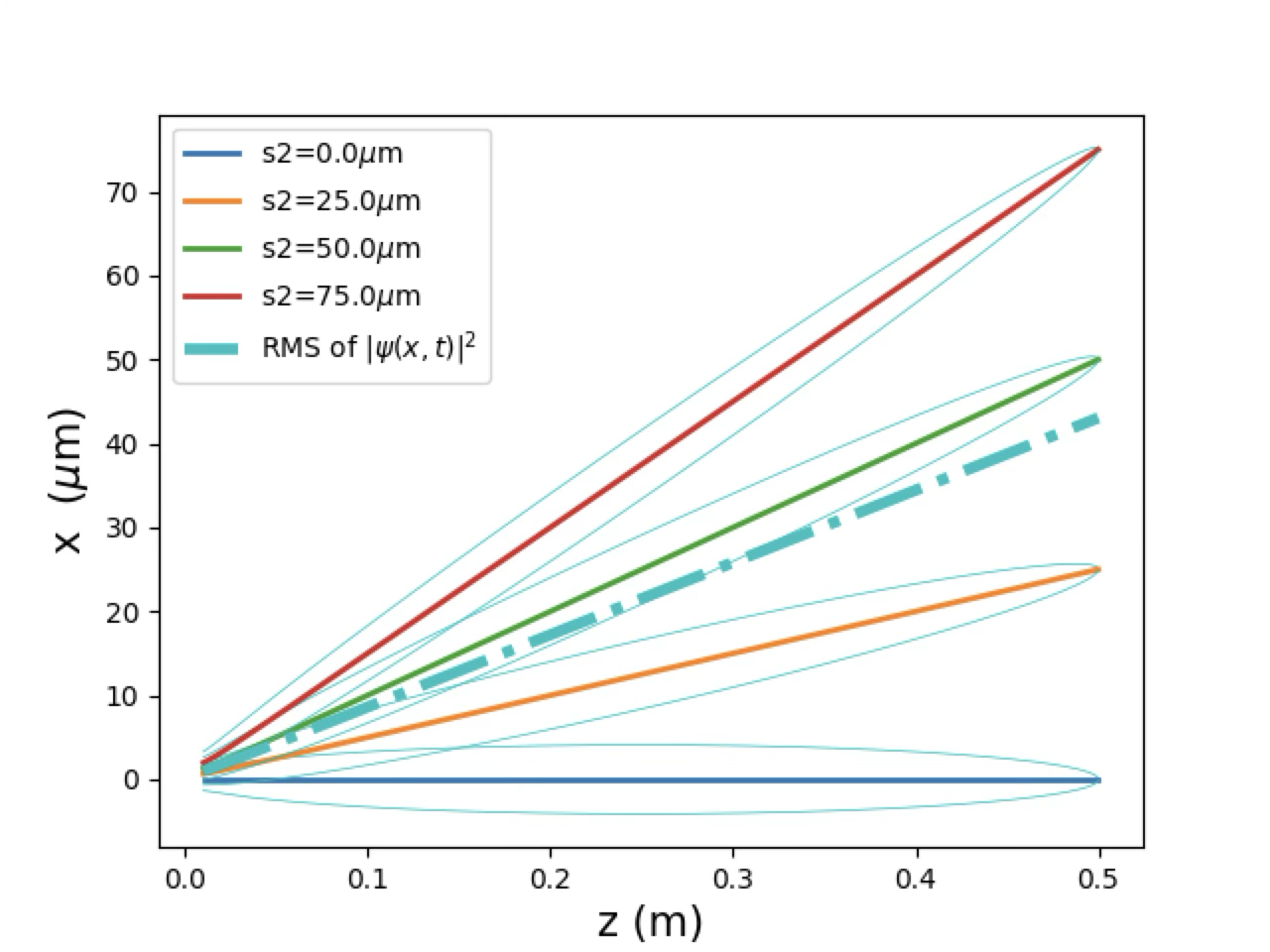}\vspace{1.0em}

\caption{The contours of $\frac{\delta P}{\delta\chi(x,z)}$. The colored lines
are the centroids $x_{c0}$ for $s_{2}=0,25,$ $50,75\mu$m, respectively.
The cyan colored contours $x_{c0}\pm\frac{1}{2}x_{\pi}$ represent
the contours of the main peaks of TPPF for various $s_{2}$. $x_{\pi}$
here is the distance from the centroid $x_{c0}$ to the point with
a phase shift from the centroid $x_{c0}$ by $\pi$. For $s_{2}=50\mu$m,
this contour corresponds to the red colored region in Fig.\ref{fig:dPdx_color}(a).
As a comparison, the RMS of the wave function $|\psi(x,t)|^{2}$ is
the thick dashed cyan line showing its width continues to spread till
the end.\label{fig:The-contours-of}}
\end{figure*}

In the example in Fig.\ref{fig:dPdx_color}(a), we display the TPPF
using a color scale for a case of $\lambda=0.541$nm (2.29kev x-ray)
with the setup parameters in Fig.1. The red contour in Fig.\ref{fig:dPdx_color}(a)
is also plotted as the contour of main peak for the case of $s_{2}=50\mu$m
in Fig.\ref{fig:The-contours-of}, where we show several different
contours of the main peaks for $s_{2}=0,25,50,75\mu$m respectively,
to show the different realizations of the ensemble represented by
the Schr{\"{o}}dinger equation solution, the wave function
$\psi(x)$ with initial condition $\psi_{1}(x_{1},z_{1})\equiv f_{1}(x_{1})$.

Fig.\ref{fig:dPdx_color} (b,c,d) shows several regions in Fig.\ref{fig:dPdx_color}(a)
near the exit detector slit 2  with details important for the next
micro-tomography discussion. In particular, Fig.\ref{fig:dPdx_color}(b)
shows that the evolution of the TPPF is continuous from $z=0.498$m
to $Z=0.5$m at the exit. 

To compare TPPF with $\psi(x)$, the RMS of $|\psi(x,z)|^{2}$ is
shown in Fig.\ref{fig:The-contours-of}, as the thick dashed cyan
line, showing its width continues to spread till the exit screen.
Fig.\ref{fig:The-contours-of} shows the contours of TPPF's central
peak emitted from entrance slit 1 with the shape of a spindle projected
in different directions; their width increases to a maximum in the
middle at about 0.25m from the end at 0.5 m. The width of the central
peak decreases after the maximum and finally, without a lens or focusing
device, converges into the end detector slits. When detector slit
2  is much narrower than entrance slit 1, as for the case of Fig.\ref{fig:dPdx_color},
the two sides of the central peaks are wave packets with increasing
width like the width of $\psi(x)$ until very close to the exit detector
slit 2  within a few hundreds $\mu$m, where they converge rapidly
into the detector slit 2   as shown in Fig.\ref{fig:dPdx_color}(b)
and Fig.\ref{fig:dPdx_color}(c). 

The rapid convergence is evident in Fig. \ref{fig:dPdx_color}(b);
in particular, it reveals a large effective convergence angle\textemdash even
though this is free-space propagation without any focusing. This behavior
highlights the influence of the exit-slit boundary condition on the
propagation of a photon wave packet: the presence of the exit slit,
which ultimately transmits the energy $h\nu$, significantly affects
the evolution of the wave packet as it approaches the slit.

In the analysis of the 1D Schr{\"{o}}dinger equation $i\frac{\partial}{\partial z}\psi=-\frac{1}{2k}\frac{\partial^{2}}{\partial x^{2}}\psi$,
we use the transformation $\phi(x,z,t)=e^{i(kz-\omega t)}\psi(x,z)$,
to study only monochromatic photons. When we introduce energy bandwidth
and study the pulse structure of the wave packet, we should be able
to analyze the time-dependent behavior of the photon wave packet and
further explore how the exit slit influences the convergence process.
This would be an important issue to be studied.

We now discuss the relation and difference between the two functions
$\psi(x,z)$ and the TPPF $\frac{\delta P_{2b}}{\delta\chi(x,z)}$.
 Fig. \ref{fig:(b)-The-pond-wave}(b) shows $\psi(x,z)$ is more like
a water wavefront propagating in a pond while the different wave packets
$\frac{\delta P_{2b}}{\delta\chi(x,z)}$ of various $s_{2}$ are more
like projectiles as shown in Fig. \ref{fig:The-contours-of}. These
projectiles, shooting into different directions, have detailed phase
information as visible fringes illustrated in Fig. \ref{fig:dPdx_color}(c)
and (d). This observation leads to the conclusion that $\psi(x)$
represents a statistical distribution of the ensemble, $\frac{1}{2P_{2b}}\frac{\delta P_{2b}}{\delta\chi(x,z)}$
represents an individual realization of the ensemble. The main point
is that the evolution of TPPF in Fig.2(b) further clarified the answer
to a fundamental question to be addressed later in Section 7: we understand
that it represents a continuous evolution of the energy distribution
$h\nu$, unlike the sudden discontinuous collapse of the probability
amplitude $\psi(x,z)$ at the exit slit.

\begin{comment}
\#\#\#!! need to replace fig.2,3 to x-ray example with same wavelength
for only one example. I should mention the resolution detemined by
slit size, not wavelength, x-ray wavelength only effect the penetration
depth, but not the resolution.
\end{comment}

\subsection*{2.3 Explore possible application to microscopy, envelope width $\sigma_{w}$,
first phase $\pi$ shift fringe spacing $x_{\pi}$, and number of
fringes $n_{f}$ within $\sigma_{w}$.}

Further examination of the details of Fig. \ref{fig:dPdx_color}(a)
indicates that the high-resolution fine fringes with significant amplitude
become hardly visible near the end detector slit 2  because when the
pin position $z$ becomes close to the detector slit 2  of narrow
width $\sigma_{2}$, the fringe spacing becomes too narrow to be visible
and needs a detailed plot. Fig. \ref{fig:dPdx_color}(d) shows that
when $L-z=0.5$mm, even at $60\mu$m, i.e.,$10\mu$m away from the
centroid $x_{c0}=50\mu$m, the amplitude of $\frac{\delta P_{2b}}{\delta\chi(x,z)}$
oscillation is still at $\pm30\text{m}^{-1}$, almost the same as
the peak value at $x_{c0}$. These details indicate a possible application
of TPPF in microscopy. To explore further, we consider the x-ray wavelength
at $\lambda=0.541$ nm, and apply Eq.(\ref{TPPF}) to a finite pin
width of 3 nm. The relative counting rate is given by the convolution
of the TPPF with the pin function, which is assumed to have a Gaussian
profile, $\Delta\chi(x-x_{p})=-\exp(-\frac{1}{2\sigma_{\chi}^{2}}\left(x-x_{p}\right)^{2})$.
Since TPPF is the functional derivative of the counting rate $P_{2b}$
with respect to the perturbation $\Delta\chi(x-x_{p})$, we have the
counting rate change as

\begin{align}
 & \frac{\Delta P_{2b}(x)}{P_{2b}}=\frac{1}{P_{2b}}\int\frac{\delta P_{2b}}{\delta\chi(x_{p},z)}\Delta\chi(x-x_{p})dx_{p}=-\left(\frac{-2\alpha_{\chi}\sigma_{\chi}^{2}}{1-2\alpha_{\chi}\sigma_{\chi}^{2}}\right)^{\frac{1}{2}}\exp\left(\frac{\alpha_{\chi}x^{2}}{1-2\alpha_{\chi}\sigma_{\chi}^{2}}\right)+c.c.\mathcal{}\label{DelPoP}
\end{align}

Here, the left-hand side of Eq.(\ref{DelPoP}) expresses the general
convolution relation defining the TPPF response to an arbitrary perturbation
$\Delta\chi$, while the right-hand side gives its explicit evaluation
for a Gaussian pin profile, which is used for numerical illustration
and for plotting Fig. 4. When the pin is sufficiently narrow, it is
equivalent to a hard-edged pin of width $\sqrt{2\pi}\sigma_{\chi}=3$nm
in this case. Since $\frac{\delta P_{2b}}{\delta\chi(x,z)}$ and $\Delta\chi(x-x_{p})$
are Gaussian, the integral is Gaussian. 

\begin{figure*}
\includegraphics[viewport=0bp 0bp 720.635bp 359.574bp,clip,scale=0.4]{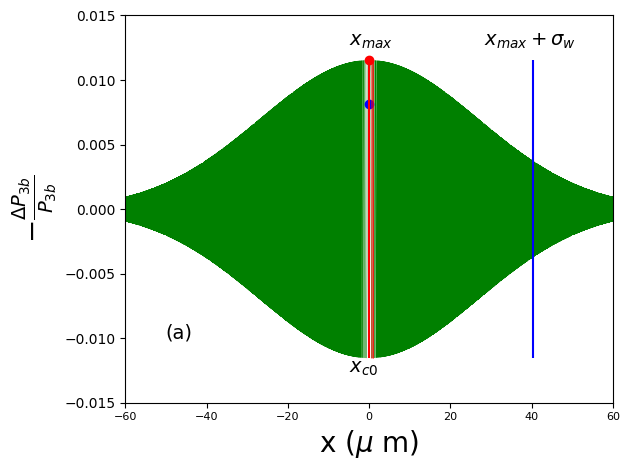}\vspace{-1.0em}\includegraphics[viewport=0bp 0bp 720.635bp 359.574bp,clip,scale=0.4]{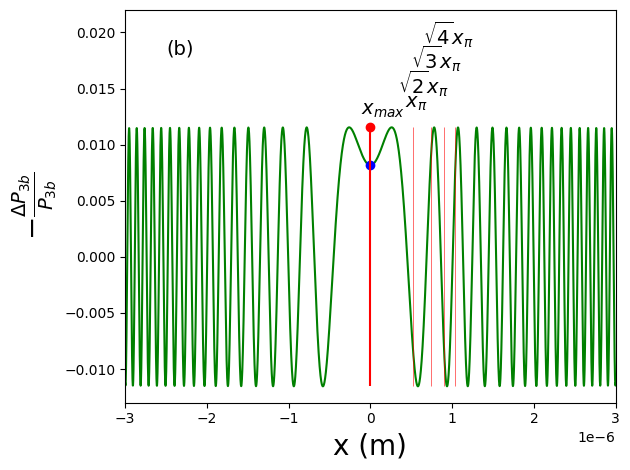}\vspace{-1.0em}

\includegraphics[viewport=0bp 0bp 719.493bp 359.729bp,clip,scale=0.5]{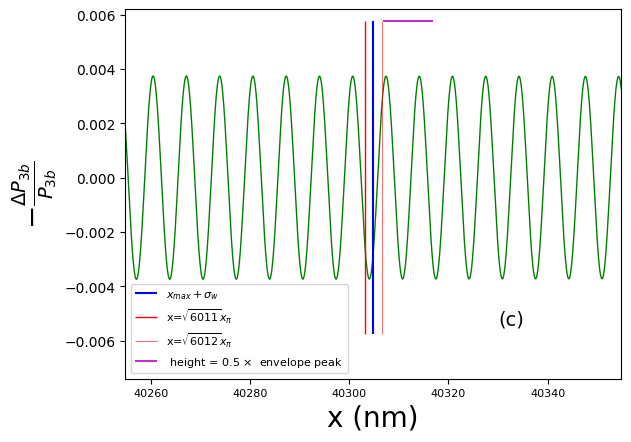}\vspace{-1.0em}

\caption{$\lambda=0.541$nm, $\sigma_{2}=0.8$ nm,$,\sigma=0.5\mu$m$,L=0.5$m,
$L-z=0.5$mm, $\Delta\chi\Delta x=3$nm (a) around peak at $|x|<60\mu$m.
(b) around peak at $|x|<3\mu$m. (c) in region around $x=x_{max}+\sigma_{w}$,
$x_{\pi}=0.52\mu$m, $\sqrt{6012}x_{\pi}-\sqrt{6011}x_{\pi}=3.35$nm
, $\sigma_{w}\approx40\mu$m. $P_{2b}=1.86\times10^{-5}$ for this
configuration and at 2.29 keV. \label{fig:nm-(a)-around}}
\vspace{-1.0em}
\end{figure*}

In Fig. \ref{fig:nm-(a)-around}, we plot the $x$ profile of $-\frac{\Delta P_{2b}(x_{p})}{P_{2b}}$
for the parameters in Fig. \ref{fig:(a)-setup}(a), that is, we choose
$L-z=0.5$mm, i.e., the pin is at 0.5mm from the detector slit 2,
and $\sigma_{2}=0.8n$m (equivalent hard-edge is $\sqrt{2\pi}\sigma_{2}=2$
nm ) , $\sigma_{1}=0.5\mu$m, $L=0.5$m, $\sqrt{2\pi}\sigma_{\chi}=3$nm.
When the pin blocks the wave function, $\Delta P_{2b}(x_{p})<0$,
the counting rate drops. Since we only limit to the case of $s_{1}=s_{2}=0$,
$x_{c}=0$, we have $\frac{\delta P_{2b}}{\delta\chi(x,z)}=P_{2b}\frac{\sqrt{-\alpha_{\chi}}}{\sqrt{\pi}}\exp\left({\displaystyle \alpha_{\chi}x^{2}}\right)+c.c.$,
and $x_{c0}=0$. Then, TPPF is simpler, and determined by the main
parameter $\alpha_{\chi}$, its amplitude and phase dominated by the
factor $\exp(\alpha_{\chi r}x^{2})=\exp(\alpha_{\chi r}x^{2})\exp(i\alpha_{\chi i}x^{2})$.
In general, we are interested in a region in Fig.1 very close to detector
slit 2 , such that $\mu,\rho,1-\xi\ll1$. For the example in Fig.
\ref{fig:nm-(a)-around}, $\mu\equiv\frac{4\pi\sigma_{1}^{2}}{\lambda L}=\frac{4\pi\times\left(0.5\mu\text{m}\right)^{2}}{0.541\text{nm}\times0.5m}=0.0116,\rho\equiv\frac{\sigma_{2}^{2}}{\sigma_{1}^{2}}=\left(\frac{0.8\text{nm}}{0.5\mu\text{m}}\right)^{2}=$$2.56\times10^{-6}$,$z_{2}=L,z_{1}=0,$
,$1-\xi=\frac{L-z}{L}=\frac{0.5\text{mm}}{0.5m}=10^{-3}$ (see Section
2.2). The approximation of $\alpha_{\chi r},\alpha_{\chi i}$ given
in Appendix III shows $\alpha_{\chi}=\alpha_{\chi r}+i\alpha_{\chi i}$
is a complex-valued parameter 

\begin{align}
 & \alpha_{\chi i}\approx\frac{\pi}{\lambda L}\frac{{\displaystyle 1}}{{\displaystyle 1-\xi}}\approx\frac{\pi\times1000}{0.541\text{nm}\times0.5m}\approx11.6\times10^{12}\text{m}^{-2}\nonumber \\
 & \alpha_{\chi r}\approx-\left(\frac{\pi}{\lambda L}2\sigma_{1}\right)^{2}\left(1+\frac{\rho}{{\displaystyle 2\left(\xi-1\right)}^{2}}\right)\approx-\left(\frac{2\pi\times0.5\times10^{-6}m}{0.541\text{nm}\times0.5m}\right)\left(1+\frac{2.56\times10^{-6}}{2\times10^{-6}}\right)\approx-3.07\times10^{8}\text{m}^{-2}\label{alphari_numrical-1}
\end{align}

This is simple and easy for back-of-the-envelope estimation with negligible
errors. The factor $\exp(i\alpha_{\chi i}x^{2})$ with $|\alpha_{\chi i}|\gg|\alpha_{\chi r}|$
and $\alpha_{\chi r}<0$, shows TPPF in Eq.(\ref{TPPF}) gives a wave
with high spatial frequency $k_{x}=2\alpha_{\chi i}x$ chirped linearly
increasing with $x$, while the amplitude factor $\exp(\alpha_{\chi r}x^{2})$
gives a slow exponential drop with increasing $x$, and reaches the
beam waist at $\sigma_{w}\equiv\sqrt{-\ensuremath{\frac{1}{2\alpha_{\chi r}}}}\approx$$\sqrt{\ensuremath{\frac{1}{2\times3.07\times10^{8}\text{m}^{-2}}}}\approx40\mu\text{m}$
determined by $\exp(\alpha_{\chi r}\sigma_{w}^{2})=\exp(-0.5)\approx0.6$.
We present these numerical parameters because we will use them later
in a practical example.

Fig. \ref{fig:nm-(a)-around}(a) shows around the center at $x=x_{c0}=0$,
where $x_{c0}$ is defined as the point of the stationary phase, the
solution of $\frac{\partial}{\partial x}\text{Im}(\alpha_{\chi}\left(x-x_{c}\right)^{2})=0$.
The phase is stationary at the origin, the phase advance increases
as $x$, the frequency increases linearly with $x$, and reaches such
a high frequency in only a few $\mu\text{m}$ that we can see only
the envelope determined by absolute value $|P_{2b}\frac{\sqrt{-\alpha_{\chi}}}{\sqrt{\pi}}\exp(\alpha_{\chi}x^{2})|$.
Its width $\sigma_{w}\approx{\displaystyle 40}\mu\text{m}$ is indicated
by the blue lines in Fig. \ref{fig:nm-(a)-around}(a). 

More details at the central peak are shown in Fig. \ref{fig:nm-(a)-around}(b),
where we see that the phase of the stationary point $x_{c0}$ is not
zero. The red lines indicates the points where the phase shift is
$\pi,2\pi,..,4\pi,..$., their spacing is not uniform, they are located
at distance from $x_{c0}$ by $x_{\pi},\sqrt{2}x_{\pi},\sqrt{3}x_{\pi},\sqrt{4}x_{\pi},..\sqrt{n_{f}}x_{\pi},..$.,
where $x_{\pi}=\sqrt{\frac{\pi}{\alpha_{\chi i}}}=0.52\mu\text{m}$.
Thus, the fringe spacing becomes much narrower at the width of $x=\sigma_{w}$. 

Fig. \ref{fig:nm-(a)-around}(c) shows the fringes at the envelope
width $\sigma_{w}$, where the blue line indicates the position at
the width $\sigma_{w}\approx40\mu\text{m}$, the two red lines indicate
the spacing between two points of index $n_{f}=10473$ and $n_{f}=10474$
is $\left(\sqrt{n_{f}+1}-\sqrt{n_{f}}\right)x_{\pi}=\sqrt{6012}x_{\pi}-\sqrt{6011}x_{\pi}=3.35\text{nm}$.
The number of fringes within the width $\sigma_{w}$ is $n_{f}=\frac{\sigma_{w}^{2}}{x_{\pi}^{2}}\approx6011$,
the fringes are so densely packed that Fig. \ref{fig:nm-(a)-around}(a)
can only show the envelope of $\frac{\Delta P_{2b}(x)}{P_{2b}}$ while
the amplitude only drops from $0.0115$ to $0.003$ for $\sqrt{2\pi}\sigma_{\chi}=3$$\text{nm}$.
If the pin is much thinner than 3nm, which is nearly half the period
in this case, at $x=\sigma_{w}$, the amplitude is $\exp(-\alpha_{\chi r}\sigma_{w}^{2})\approx0.6$
of the peak. The frequency of the waveform in a small range of Fig.
\ref{fig:nm-(a)-around}(c) is almost constant, like a sinusoidal
wave, while from the centroid at $x=0$ to the width $\sigma_{w}$,
the frequency chirps from low to high. So when $n$ is large, the
TPPF covers a larger bandwidth range. The central peak width is approximately
given by $\pm\frac{1}{2}x_{\pi}$, as illustrated in Fig. \ref{fig:nm-(a)-around}(b).
and used to indicate the contours of the main peaks in Fig. \ref{fig:The-contours-of}

\begin{comment}
to check fig.3 check with 46.2.2 profile of $-\frac{\Delta P_{3b}}{P_{3b}}$
vs. x of ``wave\_46\_xray\_2.29keV-tomography.ipynb''
\end{comment}

\begin{figure*}
\includegraphics[width=0.41\paperwidth]{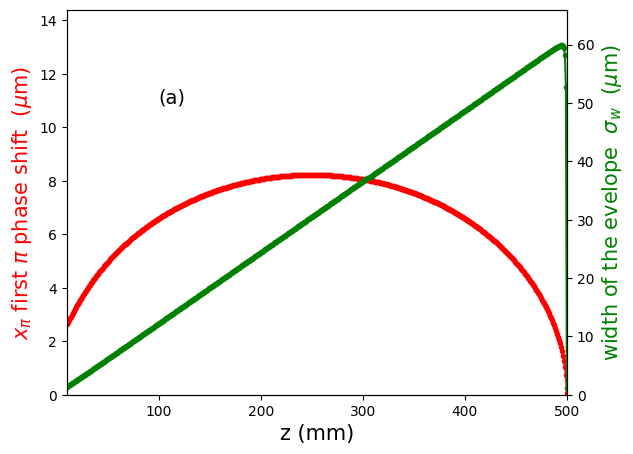}\vspace{0.0em}\includegraphics[width=0.2\paperwidth]{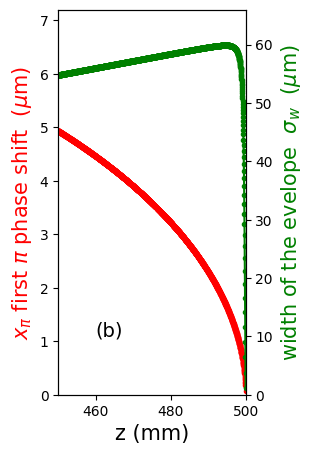}\vspace{0.0em}\includegraphics[width=0.2\paperwidth]{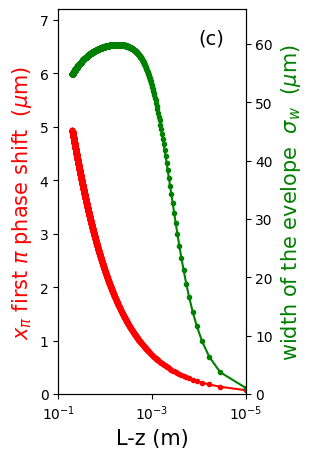}\vspace{0.0em}

\caption{Width $x_{\pi},\sigma_{w}$ for the case of $\lambda=0.541$ nm, $\sigma_{1}=0.5\mu$m$,\sigma_{2}=0.8$nm,
$s_{1}=0,$ $z_{2}=L=0.5$ m. The maximum width $x_{\pi}$ of phase
shift $\pi$ is at the midpoint $z=0.25$ m. Fig. \ref{fig:Width--for}(a):
within the valid region of $x_{\pi}=\sqrt{\frac{\pi}{\alpha_{\chi i}}}$
for $0.01m<z<0.5m-10\mu\text{m}$. When $z$ close to $0,$ the paraxial
approximation is invalid. When $z$ is too close to $0.5$ m, $\alpha_{\chi i}=0,$
$x_{\pi}$ does not exist. Fig. \ref{fig:Width--for}(b) : $\sigma_{w}=\sqrt{-\ensuremath{\frac{1}{2\alpha_{\chi r}}}}$
as the function of $z$ (the green curve) is continuous as it converges
to the exit slit near $z=L=0.5$ m as shown with details near detector
slit 2  for $50\text{mm}<L-z<10\mu\text{m}$, as compared with (a).
Fig. \ref{fig:Width--for}(c): same plot as Fig. \ref{fig:Width--for}(b)
except $z$-axis is replaced by a log scale of $L-z$, to see how
fast the TPPF converges into the exit slit.\label{fig:Width--for}}
\vspace{-1.0em}
\end{figure*}

In Fig. \ref{fig:Width--for}, we plot $\sigma_{w}$ and $x_{\pi}$
as functions of $z$, using the parameters in Fig. (\ref{fig:dPdx_color}),
for a fixed ratio $\rho\equiv\frac{\sigma_{2}^{2}}{\sigma_{1}^{2}}=2.56\times10^{-6}$.
The maximum width of $x_{\pi}$ is in the middle between the two slits,
while the maximum envelope width is much closer to the detector slit
2  when $\sigma_{2}\ll\sigma_{1}$, so close that we need to use Figs
4(b,c) to see the maximum width $\sigma_{w}$ at $L-z\approx0.5$mm.
These two plots demonstrate the rapid convergence of the wave packet
into the exit slit again, like we emphasized in Section 2.2 regarding
Fig. \ref{fig:dPdx_color}(b).

\subsection*{2.4 TPPF modulation by displacement of a 50-line comb made of 50
pins of 3nm width}

Take a point in Fig. 4(c) at $x=\sigma_{w}\approx40\mu$m as the pin
position, and take a width $\Delta x=3\text{nm}$. The probability
(we take as the counting rate) $P_{2b}=1.86\times10^{-5}$. We first
write TPPF in the form of $\frac{1}{P_{2b}}\frac{\delta P_{2b}}{\delta\chi(x,y,z)}\approx m_{TPPF}\cos(k_{x}x+\phi)$
to estimate its effect, Eq. (\ref{TPPF}). ($\phi$ is a slowly varying
phase). The approximate simple formula in Appendix III for numerical
$\alpha_{\chi r},\alpha_{\chi i}$ gives a quick estimation in Eq.(\ref{alphari_numrical-1}),
the peak modulation amplitude of the TPPF is $m_{TPPF}=|\frac{1}{P_{2b}}\frac{\delta P_{2b}}{\delta\chi(x,y,z)}|_{p}=2|\frac{\sqrt{-\alpha_{\chi}}}{\sqrt{\pi}}|\exp\left({\displaystyle \alpha_{\chi r}x^{2}}\right)_{x=\sigma_{w}}\approx2\sqrt{\frac{11.6\times10^{12}\text{m}^{-2}}{\pi}}\times0.6=2.3\times10^{6}\text{m}^{-1}$at
beam waist $\sigma_{w}$. The local phase advance rate of the phase
factor of $\frac{1}{P_{2b}}\frac{\delta P_{2b}}{\delta\chi(x,y,z)}$
, i.e., $\exp\left({\displaystyle i\alpha_{\chi i}x^{2}}\right)$
gives $k_{x}=2\alpha_{\chi i}x=2\pi/\lambda_{fringe}$, with $\lambda_{fringe}=6.7$
nm at the TPPF waist $x=\sigma_{w}$. For the opaque pin, $\Delta\chi=-1$
within the effective blocking width $\Delta x_{effective}=|\Delta\chi|\Delta x=3$
nm, Eq.(\ref{DelPoP}) gives the peak-to-peak counting rate variation
$|\frac{\Delta P_{3b}}{P_{3b}}|_{p}\approx2m_{TPPF}\Delta x_{effective}\approx0.014$
as the estimation, ignoring the width of the pin at the peak position.

The width of 3 nm is almost half of the period. The more accurate
calculation by Eq.(\ref{DelPoP}), taking into account this finite
width, gives the contribution of $\Delta x$ to the final probability
$|\frac{\Delta P_{3b}}{P_{3b}}|_{p}\approx0.0074$. If we move the
pin from a positive peak to a negative, the counting rate will change
$\pm0.0037$. The period in Fig. \ref{fig:nm-(a)-around}(c) is $T=\lambda_{fringe}\approx6.7$
nm. If we have 50 pins separated by $6.7$ nm and move in the $x$
direction, then when the pin position from the positive peak moves
by $3.35$ nm, with a $\pi$-phase shift, the intensity will change
from $18.5\%$ to $-18.5\%$. If we further increase the number of
pins, the intensity variation will increase. However, if the number
of pins is too large, the high attenuation violates the perturbation
requirement of the experiment, and the increase will not be linear.
If producing a structure 50 pins with a $6.7$nm period is difficult,
an alternative is to use some well-known structure, such as a crystal,
for a calibration of TPPF by orienting it properly to reach the $6.7$nm
period in the $x$-direction. 

\subsection*{2.5 Counting number and relation to displacement measurement error
bar for a 50-line comb of Multilayer Laue Lens (MLL) sample }

To move from an idealized 50-pin comb model to a physical realization
of the picometer displacement detection, we consider a nanostructured
multilayer (effectively a superlattice). The sample consists of 50
alternating layers of $\text{Au}$ and $\text{S\ensuremath{i_{3}}}\text{N}_{4}$
with a 6.7 nm pitch, providing the necessary high-frequency lateral
modulation. To maximize the modulation contrast at 2.29 keV, the structure
is sectioned to an optimized interaction depth of 0.56 \textgreek{m}m
along the optical axis as follows.

The technical feasibility of fabricating such periodic nanostructures\textemdash including
the deposition of sub-5 nm layers and subsequent cross-sectioning
to sub-micron depths\textemdash is well-established in the field of
Multilayer Laue Lens (MLL) development \cite{Kang,Conley,Yan}. The
gold MLL sample will provide the TPPF signal to achieve the 200 pm
displacement sensitivity of the system.

According to the Mass attenuation coefficient $\mu/\rho$ table \cite{Mass_attenuation_coefficient_table},
for gold with x-ray energy at 2.29 keV wavelength ($\lambda=0.541$
nm), we have $\mu/\rho=2389cm{{}^2}/g$ , and $\rho=19.3g/cm{{}^3}$,
so $\mu_{gold}=4.61\mu\text{m}^{-1}$. For silicon nitride, from the
CXRO/Henke database\cite{Henke} %
\begin{comment}
from CXRO/Henke database, use https://henke.lbl.gov/optical\_constants/atten2.html,
\end{comment}
, the attenuation length $L=2.2\mu\text{m}$, so $\mu_{SiN}\approx0.45\mu\text{m}^{-1}$.
With an MLL formed comb of thickness $\Delta z$, the maximum transmission
of silicon nitride is $I_{max}=\exp(-\mu_{SiN}\Delta z),$ the minimum
transmission of gold is $I_{min}=\exp(-\mu_{gold}\Delta z)$. So the
transmission is a square wave of amplitude $I_{max}-I_{min}$ with
an additional DC term of $I_{min}$, the Fourier expansion is

\begin{equation}
\Delta\chi(x-x_{p})=I_{min}+(I_{max}-I_{min})\frac{2}{\pi}\sin\left(k_{x}x\right)+...=m_{gold}\cos(k_{x}x-k_{x}x_{p})+DC+\text{\text{higher harmonics terms of }}k_{x}x\label{Delta_chi}
\end{equation}

where $m_{gold}=(I_{max}-I_{min})\frac{2}{\pi}$ , the modulation
period $\lambda_{fringe}=6.7$nm, and $k_{x}=\frac{2\pi}{\lambda_{fringe}}$.
We find the maximum modulation by finding the maximum of $I_{max}-I_{min}$
at $\Delta z=0.56\mu$m, $I_{max}=0.78$, $I_{min}=0.08.$ Then $m_{gold}=(I_{max}-I_{min})\frac{2}{\pi}\approx0.45$. 

Since the MLL sample width $\Delta x=330\text{ nm}\ll\sigma_{w}$,
TPPF is nearly sinusoidal. The convolution form of Eq(\ref{DelPoP})
gives the variation of detection rate, ignoring the term with fast
oscillating phase $2k_{x}x$ in the following integral, and also ignoring
the DC term and other higher harmonics terms, we have 

\begin{align}
 & \frac{\Delta P_{2b}(x_{p})}{P_{2b}}=\int\frac{1}{P_{2b}}\frac{\delta P_{2b}}{\delta\chi(x,z)}\Delta\chi(x-x_{p})dx\nonumber \\
 & =\int m_{TPPF}\cos(k_{x}x+\phi)m_{gold}\cos(k_{x}x-k_{x}x_{p})dx=\frac{1}{2}m_{TPPF}m_{gold}\cos\left(k_{x}x_{p}+\phi\right)\Delta x\label{DePoP}
\end{align}

where $\phi$ is a phase chosen such that at $x_{p}$ the phase of
TPPF is $k_{x}x_{p}+\phi$. Thus, when we move the MLL gold sample
$x_{p}$ by 3.35nm, using $m_{TPPF}=2.3\times10^{6}\text{m}^{-1}$
given above, the peak-to-peak variation is $|\frac{\Delta P_{2b}}{P_{2b}}|_{p}=m_{TPPF}m_{gold}\Delta x=2.3\times10^{6}\text{m}^{-1}\times0.45\times330\text{nm}\approx0.34$
for the setup parameters in Fig. 4(c), idealized with complete attenuation
outside the aperture.

While the physical width of the gold comb ($\Delta x=330\text{ nm}$)
is wider than the target displacement sensitivity, it remains significantly
smaller than the $1\ \mu\text{m}$ FWHM of the scanning resonance
peak. Calculations of the effective interaction width confirm that
the sinusoidal TPPF approximation remains highly accurate, maintaining
an effective-to-physical width ratio of 0.996 in the lower-energy
case and 0.97 in the higher-energy case. Thus, the finite width of
the comb does not induce non-linear phase blurring or compromise the
target precision.

To get sufficient accuracy, we use the Poisson distribution formula
$P(k)={\displaystyle \frac{n^{k}e^{-n}}{k!}}$, its RMS is $\sqrt{n}$
where $n$ is the mean rate of events during a fixed interval. The
shot noise is $\delta N_{2}=\sqrt{N_{2}}$ , it leads to an effective
displacement error $\delta x_{p}$ such that $\delta N_{2}=\delta\Delta N_{2}=N_{2}\delta\left(\frac{\Delta P_{2b}}{P_{2b}}\right)=-\frac{1}{2}k_{x}N_{2}m_{TPPF}m_{gold}\sin\left(k_{x}x_{p}+\phi\right)\Delta x\delta x_{p}$.
At the most sensitive phase $k_{x}x_{p}+\phi=-\pi/2$, this reaches
maximum with peak-to-peak value of $\delta N_{2}=\sqrt{N_{2}}=k_{x}N_{2}\left(m_{TPPF}m_{gold}\Delta x\right)\delta x_{p}=k_{x}N_{2}|\frac{\Delta P_{2b}}{P_{2b}}|_{p}\delta x_{p}$.
Thus $\delta x_{p}=\frac{\sqrt{N_{2}}}{k_{x}N_{2}|\frac{\Delta P_{2b}}{P_{2b}}|_{p}}=\frac{\lambda_{fringe}}{2\pi\sqrt{N_{2}}|\frac{\Delta P_{2b}}{P_{2b}}|_{p}}$.
If we choose the sensitivity to be $\delta x_{p}=200$ pm, let $N_{2}$
denote the required incident photon numbers, we have

\begin{align}
 & N_{2}=\left(\frac{\lambda_{fringe}}{2\pi\delta x_{p}|\frac{\Delta P_{2b}}{P_{2b}}|_{p}}\right)^{2}=\left(\frac{6.7\text{nm}}{2\pi\times200pm\times0.34}\right)^{2}=246\label{N2a}
\end{align}

\begin{comment}
see 40.8 of no\_bath\_free\_space.lyx
\end{comment}

Since $P_{2b}^{(0)}=1.86\times10^{-5}$ for Fig. \ref{fig:nm-(a)-around},
$N_{10}=\frac{N_{2}}{P_{2b}^{(0)}}=\frac{246}{1.86\times10^{-5}}=1.32\times10^{7}$
is the required incident photon number to reach the sensitivity $\delta x_{p}=200$
pm, both of which assume the idealized detector slit 2  geometry. 

The signal-to-noise ratio SNR is 

\begin{align}
 & S_{n}\equiv\frac{\Delta N_{2}}{\delta N_{2}}=|\frac{\frac{\Delta N_{2}}{N_{2}}}{\frac{\delta N_{2}}{N_{2}}}|_{p}=\frac{|\frac{\Delta P_{2b}(s)}{P_{2b}}|_{p}N_{2}}{\sqrt{N_{2}}}=|\frac{\Delta P_{2b}(s)}{P_{2b}}|_{p}\sqrt{N_{2a}}=0.34\times\sqrt{246}=5.3\label{SNR}
\end{align}

The error bar is $1/S_{n}=\frac{\delta N_{2a}}{\Delta N_{2a}(x=\sigma_{w})}=19\%$.
Increasing the photon number to $1.32\times10^{9}$ will reduce the
error to $1.9\%$.

We remark that while a $\delta$-function kernel in Eq. (\ref{DePoP})
would yield a standard Radon transform, our sinusoidal kernel produces
a direct physical Fourier transform, as we shall explore in the following
section. By sampling frequency components directly, this configuration
performs the requisite transformation for the Fourier Slice Theorem
physically. Specifically, by replacing the gold modulation $m_{gold}$
with the Fourier coefficient $m_{sample}$ of the sample $\Delta\chi(x-x_{p})$,
Eqs.(\ref{DePoP},\ref{Delta_chi}) provide the sample\textquoteright s
complex-valued structure as analyzed in \cite{yu-1}, where $m_{sample}$
gives the Fourier coefficient amplitude and $k_{x}x_{p}$ gives its
phase. This reduces reliance on iterative phase-retrieval procedures
in the present detection scheme and provides a basis for phase-encoded
frequency-domain analysis, suggesting a pathway toward frequency-domain
tomography.

TPPF manifests as a stable, reproducible structure consistent with
an objectively existing field of high-resolution fringes. We will
explore it as a tool to probe microscopic structure.

\subsection*{2.6 Photon budget to reach 200 pm precision with a detector behind
a practical 2 nm wide MML slit for the MLL sample}

\begin{figure*}
\includegraphics[width=0.41\paperwidth]{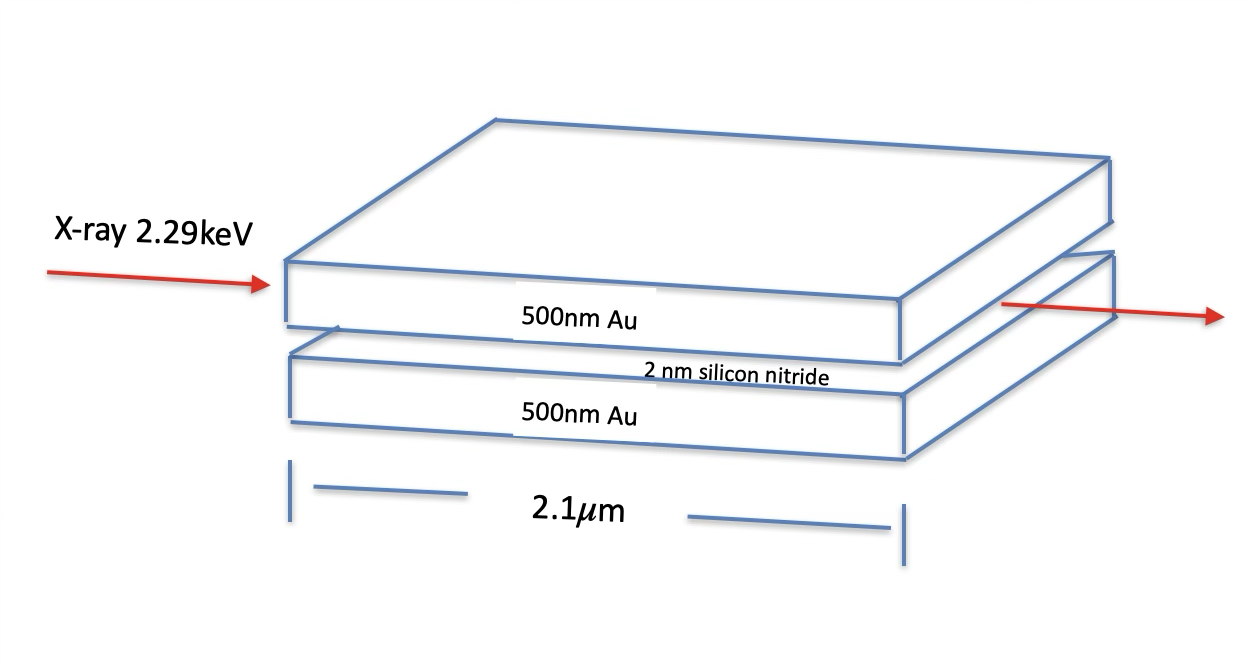}\vspace{0.0em}

\caption{Illustration of a 2 nm MLL slit \label{MMLslit}}
\vspace{-1.0em}
\end{figure*}

We need to consider replacing the idealized 2 nm detector slit by
a practical slit. The MLL technology \cite{Kang,Conley,Yan} used
for the MLL sample in the last section can be applied for this slit,
by using only one 2 nm layer of silicon nitride between two layers
of 100 to 500 nm gold, as illustrated in Fig.(\ref{MMLslit}).

The transmission of 2.1$\mu$m gold is $\exp(-4.61\times2.1)=6\times10^{-5}$.
For the silicon nitride 2.1 $\mu$m core the transmission is $I_{SiN}=\exp(-0.45\times2.1)=0.39$.
So the MLL slit of 2 nm width, 2.1 $\mu\text{m}$ deep, will have
$P_{2b}=I_{SiN}P_{2b}^{(0)}$, where $P_{2b}^{(0)}$ is the detection
probability (counting rate) for an idealized detector slit 2  with
100\% transmission. The detector slit 2  profile in Eq.(\ref{complex_dpdchi})
is $f_{2}^{2}(x_{2})=I_{SiN}\left(f_{2}^{(1)}(x_{2},\sigma_{2}^{\{0\}})\right)^{2}$
where the hard-edged profile $f_{2}^{(1)}(x_{2},\sigma_{2}^{\{0\}})=1$
within $|x_{2}|\le1$nm for $\sqrt{2\pi}\sigma_{2}^{\{0\}}=2$ nm
can be approximated by a Gaussian profile of RMS size $\sigma_{2}^{\{0\}}$.
Eq.(\ref{complex_dpdchi}) shows $\frac{\delta P_{2b}^{(c)}}{\delta\chi(x,z)}$
and $P_{2b}$ are proportional to $I_{SiN}$, hence $P_{2b}=I_{SiN}P_{2b}^{(0)}$.
Thus, $\frac{1}{P_{2b}}\frac{\delta P_{2b}^{(c)}}{\delta\chi(x,z)}$
is independent of $I_{SiN}$. That is, $\frac{\Delta P_{2b}(x_{p})}{P_{2b}}$
in Eq.(\ref{DePoP}) remains the same when the detector slit 2  is
replaced by the MLL detector slit. Hence, the photon count $N_{2}=246$
remains the same. The required photon number to achieve precision
$\delta x_{p}=200$pm becomes $N_{1}=\frac{N_{2}}{P_{2b}}=\frac{N_{2}}{I_{SiN}P_{2b}^{(0)}}=\frac{246}{0.39\times1.86\times10^{-5}}=3.4\times10^{7}$
for the 2 nm MLL slit of $2.1\mu\text{m}$ thick and the MML gold
sample. We note that while this lower-energy configuration ($2.29\text{ keV}$)
requires $3.4\times10^{7}$ incident photons, the higher-energy example
($6\text{ keV}$) analyzed in Section 4 scales to a more efficient
total photon budget of approximately $1\times10^{7}$.

While Eq. (\ref{P2b_TPPF_relation}) is derived under the weak-phase
approximation, this framework remains robust for highly refracting
objects; as shown in Appendix IV, the lumped system transfer function
$C_{\text{bulk}}$\textemdash representing the collective propagation
effects of the thick medium\textemdash remains a separable factor,
allowing the picometer displacement detection model to be generalized
for highly refracting objects through experimental calibration. This
separability ensures that the Fourier\textendash Radon framework is
a universal tool for both biological and industrial X-ray physics.

\section*{3. Relation between TPPF and Radon transform}

\subsection*{3.1 TPPF is related to the Fourier transform of the Radon transform
in 3D tomography}

In the convolution form of Eq.(\ref{DelPoP}), we can replace the
perturbation of the pin represented by $\int\Delta\chi(x)dx$ with
a perturbation by a sample represented by attenuation $-\int f(x,y,z)dxdydz$
where $f(x,y,z)$ is a real function when we only consider attenuation.
If the sample causes a phase shift, it is a complex function, but
the formulation is the same.

\begin{align}
 & \frac{\Delta P_{2b}(s)}{P_{2b}}=\frac{\Delta P_{2b}^{(c)}(s)}{P_{2b}}+c.c.=-\frac{1}{P_{2b}}\int\frac{\delta P_{2b}^{(c)}}{\delta\chi(x,z)}f(x-s,y,z)dxdydz+c.c.\label{DelPoP-1}
\end{align}

Here $\frac{\delta P_{2b}^{(c)}}{\delta\chi(x,z)}$ is given in Eq.(\ref{dP2bdchi-1})
and Eq.(\ref{TPPF}). Because the y translational symmetry of Fig.
\ref{fig:(a)-setup} when the slits are long and thin, the functional
derivative of $P_{2b}$ with respect to the perturbation of a point
in $\Delta\chi(x,y,z)=f(x,y,z)$ in Eq.(\ref{pertubation:}) i.e.
the TPPF function $\frac{\delta P_{2b}}{\delta\chi(x,z)}$, is independent
of $y$, as given by Eq.(\ref{TPPF}). For an object $f(x,y,z)$ of
the order of a few $\mu\text{m}$, the $z$ dependence of $\frac{\delta P_{2b}}{\delta\chi(x,z)}$
is also negligible because its only dependence on $z$ in Eq.(\ref{TPPF})
is through the $\alpha_{\chi}$ dependence on $\xi=\frac{z}{L}$.
For the setup in the example of Fig. \ref{fig:nm-(a)-around}, $L=0.5m$,
and the distance to the detector slit 2  is $L-z=0.5$mm, so a variation
of $z$ by a few $\mu$m compared with the distance of 0.5mm is negligible.
In addition, in the following application of Eq.(\ref{DelPoP-1})
for 3D micro-tomography, we can take this tiny variation into the
algorithm without any significant differences because the explicit
analytical expression $\frac{\delta P_{2b}}{\delta\chi(x,z)}$ in
Eq.(\ref{TPPF}) is not needed; we only need the high-resolution structure
of the TPPF and its Fourier transform. In a practical experiment,
its specific value should be measured or calibrated by known samples;
there is no need to calculate it accurately. Because the frequency
of the waveform $\frac{\delta P_{2b}}{\delta\chi(x,z)}$ in a small
range within a few microns of Fig.\ref{fig:nm-(a)-around}(c) is almost
constant, like a sinusoidal wave, $\frac{\Delta P_{2b}(s)}{P_{2b}}$
is approximately a Fourier transform of $\int f(x,y,z)dxdydz$ . Once
$\frac{\delta P_{2b}}{\delta\chi(x,z)}$ is measured in an experiment,
$\frac{\delta P_{2b}^{(c)}}{\delta\chi(x,z)}$ can be calculated readily
by a Hilbert transform, as explained at the end of Section 2.1 (see
Appendix I).

In Eq.(\ref{DelPoP-1}) we introduced a translation $s$ in $x$ direction
representing a scan of the sample such that $\frac{\Delta P_{2b}(s)}{P_{2b}}$
becomes a sum of a convolution between $\frac{\delta P_{2b}^{(c)}}{\delta\chi(x,z)}$
and $f(x-s,y,z)$ and its complex conjugate. If the range of $f$
is a few microns, the range is smaller than the range of TPPF in Fig.
\ref{fig:nm-(a)-around}(a). When we change $s$, the overlap between
$\frac{\delta P_{2b}^{(c)}}{\delta\chi(x,z)}$ and $f(x-s,y,z)$ will
be in regions of different frequencies because the frequency is chirped
in the exponent $\alpha_{\chi}x^{2}$ of $\frac{\delta P_{2b}^{(c)}}{\delta\chi(x,z)}$
in Eq. (\ref{TPPF}). When $s$ increases, $\frac{\delta P_{2b}^{(c)}}{\delta\chi(x,z)}$
samples higher frequency components in $f$. Since $\frac{\delta P_{2b}^{(c)}}{\delta\chi(x,z)}$
is almost sinusoidal within a few micron range, Eq.(\ref{DelPoP-1})
is approximately a Fourier transform of $f$ at a frequency specified
by $s$. Hence, the scan of $s$ corresponds to the measurement of
the spectrum of $f$ in the x-direction. $\frac{\Delta P_{2b}(s)}{P_{2b}}$
is approximately a Fourier transform of $f(x,y,z)$ in the x-direction.

We now compare Eq.(\ref{DelPoP-1}) with the Radon transform \cite{3DRadon}
$p(s,\overrightarrow{\alpha})$ in the 3D tomography for direction
$\overrightarrow{\alpha}$. When we choose the sample reference coordinates
$u,v,w$ such that $u$ is in $\overrightarrow{\alpha}$ direction,
it is the projection of the sample $f(s,v,w)$ from a plane in the
sample perpendicular to $\overrightarrow{\alpha}$ at distance s from
the origin onto a line in the $\overrightarrow{\alpha}$ direction,
it converts the 3D density $f$ into a line density

\begin{align}
 & p(s,\overrightarrow{\alpha})=\int\int\int f(\overrightarrow{x})\delta(\overrightarrow{x}\cdot\overrightarrow{\alpha}-s)d\overrightarrow{x}=\int\int\int f(u,v,w)\delta(\overrightarrow{x}\cdot\hat{u}-s)dudvdw\nonumber \\
 & =\int\int\int f(u,v,w)\delta(u-s)dudvdw=\int\int f(s,v,w)dvdw\label{Radon}
\end{align}

\begin{figure*}
\includegraphics[scale=0.5]{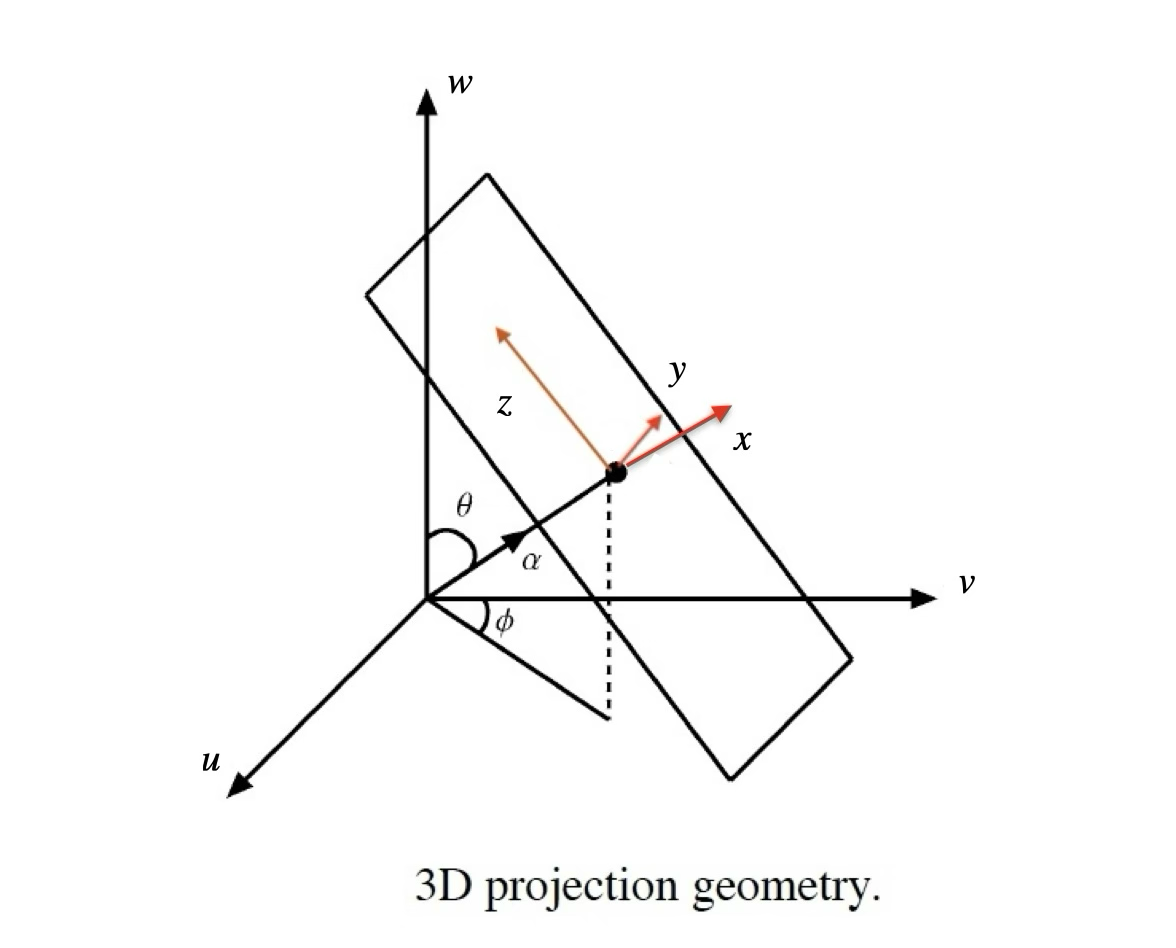}\vspace{-1.0em}

\caption{Relation between the sample reference frame $(u,v,w)$ and the experiment
setup reference frame $(x,y,z)$\label{fig:Relation-between-x-and-frequency}}
\vspace{-1.0em}
\end{figure*}

If we choose the $x$-axis of the TPPF measurement to be in $\overrightarrow{\alpha}$
direction, i.e., $\hat{x}=\overrightarrow{\alpha}$, the comparison
shows the counting rate $\frac{\Delta P_{2b}(s)}{P_{2b}}$ is approximately
the Fourier transform of the Radon transform mentioned in \cite{3DRadon},
where $\int f(x,y,z)dydz$ is the projection of $f$ in the $y,z$
plane onto the $x$-axis, so it converts $f$ into a line density
in the $x$ direction. In Fig. \ref{fig:Relation-between-x-and-frequency},
we plot the relation between the sample reference frame and the experiment
setup reference frame of Fig. \ref{fig:(a)-setup}(a). Varying the
Euler angle $\theta$ and $\phi$ will change the orientation of the
sample while the experimental setup of $x,y,z$ is fixed. The plot
only gives one specific choice of the orientation where $z$ is pointing
to the $w$-axis so that the $w$-axis is within the $\{x,z\}$ plane.
For this orientation, $\hat{x}=\overrightarrow{\alpha}=(\sin\theta\sin\varphi,\sin\theta\cos\varphi,\cos\theta)$
$,\hat{y}=(-\cos\varphi,\sin\varphi,0)$, $\hat{z}=(-\cos\theta\sin\varphi,-\cos\theta\cos\varphi,\sin\theta)$
in the $\{u,v,w\}$ coordinate system. However, any rotation of the
sample around the $x$-axis would not change the $x$-line density
and gives another choice of the scan orientation of the sample, so
our choice here is not unique.

Once the orientation is chosen, the coordinate transform gives the
Radon transform function,

\begin{align}
 & p(s,\overrightarrow{\alpha})=\int\int f(s,v,w)dvdw=\int\int f(s,y,z)dydz.\label{Radon2}
\end{align}

\begin{comment}
see 40.5.3.2 of no\_bath\_free\_space.lyx
\end{comment}

\begin{comment}
we were forced to use $\omega$ as the Fourier transfrom variable
of $x$: because $k_{x}=\frac{2\pi}{\lambda}$ has been used for the
longitudinal wave number, we could use $k_{x}$ but $\omega$ has
been used in the fourier transform of Radon transform, so we just
let $\omega=k_{x}$
\end{comment}
{} 

For the 3D tomography, the starting point is to calculate the Radon
transform $\int\int f(s,y,z)dydz$. For this, we rewrite Eq.(\ref{DelPoP-1}).
Since for practical case $s_{1}=s_{2}$, $x_{c}=0$,

\begin{align}
 & g(s)=\frac{\Delta P_{2b}}{P_{2b}}=\int G^{(c)}(x)f(x-s,y,z)dxdydz+c.c.\equiv g^{(c)}(s)+g^{(c)*}(s)\nonumber \\
 & G^{(c)}(x)\equiv-\frac{1}{P_{2b}}\frac{\delta P_{2b}^{(c)}}{\delta\chi(x,z)}=-\frac{\sqrt{-\alpha_{\chi}}}{\sqrt{\pi}}\exp(\alpha_{\chi}x^{2})\label{Gx}\\
 & \hat{g}^{(c)}(\omega)=\hat{G}^{(c)}(\omega)\hat{f}(\omega)\nonumber 
\end{align}

$g^{(c)}(s)$ here can be calculated from $g(s)$ by Hilbert transform
(See Appendix I) once it is measured, similar to a calibration of
$G^{(c)}(x)$. Since the counting rate variation scan $g^{(c)}(s)$
is the convolution of the complex TPPF $G^{(c)}(x)$ and the line
density $\int f(x,y,z)dydz$, its Fourier transform $\hat{g}^{(c)}(\omega)$
(FT) is the product of the FT $\hat{G}^{(c)}(\omega)$ and the FT
$\hat{f}(\omega)$, i.e., the Fourier transform of Radon transform
$p(s,\overrightarrow{\alpha})$. Let $\overrightarrow{k}=\omega\overrightarrow{\alpha}$,
$\overrightarrow{k}\cdot\overrightarrow{x}=\omega x$, ($\omega$
is the transverse wavenumber, i.e., we denote the spatial frequency
$k_{x}=\omega$ in the $x$ direction),

\begin{align}
 & \hat{f}(\omega)=\hat{f}(\omega,\overrightarrow{\alpha})=\hat{f}(\overrightarrow{k})=\int\int\int f(\overrightarrow{x})\exp(-\overrightarrow{ik}\cdot\overrightarrow{x})d\overrightarrow{x}\nonumber \\
 & =\int dx\exp(-i\omega x)\int\int f(x,y,z)dydz\nonumber \\
 & \hat{f}(\omega)=\int dx\exp(-i\omega x)p(x,\overrightarrow{\alpha})\label{ftilde}
\end{align}
$\hat{G}^{(c)}(\omega)$ is the Fourier transform of $G^{(c)}(x)$,

\begin{align}
 & \hat{G}^{(c)}(\omega)\equiv\int\exp(-i\omega x)G^{(c)}(x)dx\nonumber \\
 & =\int dx\exp(-i\omega x)\int\int\frac{\sqrt{-\alpha_{\chi}}}{\sqrt{\pi}}\exp(\alpha_{\chi}\left(x\right)^{2})dydz\label{Gk}
\end{align}

Thus, $\hat{f}(\omega)$ can be calculated as

\begin{align}
 & \hat{f}(\omega)=\frac{\hat{g}^{(c)}(\omega)}{\hat{G}^{(c)}(\omega)}\label{ft}
\end{align}

In short, structure function $f(x,y,z)$ can be obtained from $G(x)=-\frac{1}{P_{2b}}\frac{\delta P_{2b}}{\delta\chi(x,z)}$
measured in an experiment as a calibration, and $g(t)$ measured by
scanning the sample, according to the following steps:
\begin{itemize}
\item 1. $G(x)=-\frac{1}{P_{2b}}\frac{\delta P_{2b}}{\delta\chi(x,z)}\overset{\text{Hilbert Transform}}{\longrightarrow}G^{(c)}(x)\overset{\text{Fourier Transform }}{\longrightarrow}\hat{G}^{(c)}(\omega)$. 
\item 2. Measure by scan $g(t)\overset{\text{Hilbert Transform}}{\longrightarrow}g^{(c)}(x)\overset{\text{Fourier Transform }}{\longrightarrow}\hat{g}^{(c)}(\omega)$.
\item 3. $\hat{f}(\omega)=\frac{\hat{g}^{(c)}(\omega)}{\hat{G}^{(c)}(\omega)}$ 
\item 4.$\hat{f}(\omega)\overset{\text{Inverse Fourier Transform }}{\longrightarrow}p(x,\overrightarrow{\alpha})=\int\int f(x,y,z)dydz$
\end{itemize}
In the following, we find the bandwidth where $|\hat{G}^{(c)}(\omega)|$
is sufficiently larger than the noise of the system so $\hat{f}(\omega)=\frac{\hat{g}^{(c)}(\omega)}{\hat{G}^{(c)}(\omega)}$
can be calculated with sufficient precision, which in turn determines
the resolution of the 3D tomography. 

\subsection*{3.2 Bandwidth of $\hat{G}^{(c)}(\omega)$ and the resolution of the
3D tomograph at the bandwidth}

In the application of 3D tomography, the error in the image originates
from the measurement of $g(s)$ and $G(x)$. The error is determined
by the method of Section 2.5 by photon counting number. We may use
$\hat{G}^{(c)}(\omega)$ calculated from $G(x)$ in Eq.(\ref{Gx})
if the detector slit 2  is sufficiently narrow and uniform; otherwise,
in case the deviation from the idealized slit is significant, $\hat{G}^{(c)}(\omega)$
should be calculated from the measurement of the counting rate $G(x)$.
Our discussion on the property of TPPS following the Eq.(\ref{DelPoP-1})
shows that for each $s$ in $\frac{\Delta P_{2b}(s)}{P_{2b}}$ of
the equation, there is a corresponding $\omega$ in the Fourier transform
$\hat{G}^{(c)}(\omega)$ of $\frac{\delta P_{2b}^{(c)}}{\delta\chi(x,z)}$
that is peaked. This means that when we scan $s$ from $0$ to $\sigma_{w}$,
the $\hat{G}^{(c)}(\omega)$ has a non-zero range as its bandwidth
with a low and high frequency limit. Within this bandwidth $|\hat{G}^{(c)}(\omega)|$
is sufficiently large so that the relative error of $\frac{\Delta\hat{G}^{(c)}(\omega)}{|\hat{G}^{(c)}(\omega)|}$
is small, and the calculation of $\hat{f}(\omega)$ from the corresponding
$\hat{g}^{(c)}(\omega)$ is less sensitive to error $\delta\hat{g}^{(c)}(\omega)/\hat{G}^{(c)}(\omega)$. 

\begin{comment}
$\Delta x\approx1/(2f_{0})=\pi/\omega_{0}$ (resolution definition
cite Goodman; cite a Fourier-ptychography/microscopy paper for the
half-pitch formula).see 5.9 The relation between cutoff frequency
$f_{max}=\omega_{max}/(2\pi)$, noise (SNR=1) and resolution $\frac{1}{2f_{max}}=\frac{\pi}{\omega_{max}}$
, when $\omega=1/\text{nm}$ ,resolution is $\pi\text{nm}=3.14\text{nm}$,
in file:///Users/lihuayu/Desktop/QMwave\_collapse/no\_bath\_free\_space/n2o\_bath\_free\_space\_back2.lyx
\end{comment}

\begin{comment}
\#\#comments see 40.5.3.4 for $\hat{G}^{(c)}(\omega)$
\end{comment}

In the Fourier transform of the TPPF Eq.(\ref{Gk}), because of the
discussion in subsection 3.1, we can ignore the $y$ and $z$ dependence
in the integral when the slits are sufficiently long and narrow. Assuming
the slit length is $\Delta y$ and the sample size in the direction
of $z$ as $\Delta z$, we find $\hat{G}^{(c)}(\omega)=\Delta y\Delta z\exp\left(\frac{\omega^{2}}{4\alpha_{\chi}}\right)$.
The bandwidth $\sigma_{\omega}$ is determined as $\frac{1}{4}\left(\frac{1}{\alpha_{\chi}}\right)_{r}=-\frac{1}{2\sigma_{\omega}^{2}}$,
where $\left(\frac{1}{\alpha_{\chi}}\right)_{r}$ is the real part
of $\frac{1}{\alpha_{\chi}}$. If we cut off $\hat{G}^{(c)}(\omega)$
at $\sigma_{\omega}$ , the resolution limited by this bandwidth is
\cite{resolution} $\frac{T}{2}=\frac{\pi}{\sigma_{\omega}}=\pi\sqrt{-\frac{1}{2}\left(\frac{1}{\alpha_{\chi}}\right)_{r}}\approx\pi\sqrt{-\frac{1}{2}(-2\sigma_{2}^{2})}=\pi\sigma_{2}$
(see Appendix III for the approximation of $\alpha_{\chi}$), where
$T$ is the period at the bandwidth $\sigma_{\omega}$. We will first
take this as an estimate. At this cutoff, $\hat{G}^{(c)}(\omega)$
is reduced by a factor $e^{-\frac{1}{2}}\approx0.6$, the noise of
20\% would not significantly affect the image reconstruction. For
comparison, in the case of the example in Section 2.5, the error bar
would be $19\%$. Thus, within the bandwidth $\sigma_{\omega}$, $\hat{G}^{(c)}(\omega)$
is far from zero, and the solution $\hat{f}(\omega)=\frac{\hat{g}^{(c)}(\omega)}{\hat{G}^{(c)}(\omega)}$
is stable. If the noise of the measurement is smaller, we may cut
off at a higher frequency, or scan $s$ to beyond the TPPF width (
$s>\sigma_{w}$), and the resolution may be improved if other noise,
such as detector noise, is lower.

To minimize the resolution, we observe that $\alpha_{\chi}$ give
by Eq.(\ref{TPPF}) depends on $\mu\equiv\frac{4\pi\sigma_{1}^{2}}{\lambda L}$,
$\rho\equiv\frac{\sigma_{2}^{2}}{\sigma_{1}^{2}}$ and $z=\xi L$.

\begin{figure*}
\includegraphics[viewport=0bp 0bp 720.247bp 360.39bp,clip,scale=0.5]{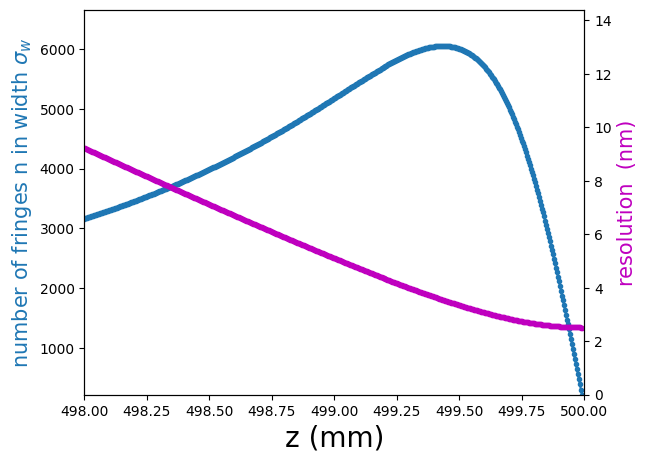}\vspace{-1.0em}

\caption{Number of fringes $n_{f}=\frac{\sigma_{w}^{2}}{x_{\pi}^{2}}$ in the
width $\sigma_{w}$ and the resolution vs. $z$. For the case of the
example in Fig. \ref{fig:nm-(a)-around}: $\lambda=0.541\text{nm},\sigma_{2}=0.8\text{nm},\sigma_{1}=0.5\mu,L=0.5m$\label{fig:resolution}}
\vspace{-1.0em}
\end{figure*}

Fig. \ref{fig:resolution} shows the resolution vs. $z$ (the magenta
curve) for the example in the setup of Fig. \ref{fig:nm-(a)-around}
, together with the number of fringes $n_{f}=\frac{\sigma_{w}^{2}}{x_{\pi}^{2}}$.
The minimum of the resolution is close to the exit slit at $L-z=0.64\mu$m.
But we choose $L-z=0.5$mm for the example of Fig.\ref{fig:nm-(a)-around}
because we also need to increase the number of fringes $n_{f}$ within
the width $\sigma_{w}$. Choosing $L-z=0.5$mm increases the $s$
scan range in the Radon Fourier transform of Eq.(\ref{DelPoP-1}).
Thus, the scan covers a larger bandwidth from large $T$ near $x=0$
to small period $T$ at $x=\sigma_{w}.$ 

If we choose $L-z<0.5$ mm, the frequency is higher according to Eq.(\ref{alphari_numrical-1})
with increased amplitude, thus the signal will increase at the expense
of a smaller distance from the detector slit and a smaller number
of fringes.%
\begin{comment}
\#comments see section 40.7 and 46.2.1 of of no\_bath\_free\_space.lyx
\end{comment}

To further understand the example of Fig. \ref{fig:Width--for} near
detector slit 2  and the relation of the resolution to $\sigma_{1},\sigma_{2},\lambda L$,
and $z$ more quantitatively, we notice that when $\mu\ll1$, $\rho\ll1$
and $\xi\approx1$ (see section 2.3), there is a region of a few mm
from the exit but not too close to it ($\xi$ is not too close to
$1$), where the parameters satisfy the condition $|\frac{\alpha_{\chi r}}{\alpha_{\chi i}}|=\frac{\pi\sigma_{2}^{2}}{{\displaystyle 2\left(1-\xi\right)\lambda L}}\ll1$
(See Appendix III for the approximation of $\alpha_{\chi}$ under
this condition, which provides a convenient expression for quick experimental
estimates.) This implies that the decay rate $|\alpha_{\chi r}|$
of the TPPF envelope with distance from the axis is much smaller than
its oscillation rate $|\alpha_{\chi i}|$, provided that \ensuremath{z}
is sufficiently close to detector slit 2 \textemdash but not too close,
as an excessively small value of $1-\xi$ would invalidate this condition.
A comparison with a hard-edged detector slit 2  shows that, under
the same conditions, a hard-edged slit of width $\sqrt{2\pi}\sigma_{2}$
is effectively equivalent to a Gaussian slit of width $\sigma_{2}$,
with negligible difference in the resulting TPPF near beam waist at
$x\approx\sigma_{w}$.

Thus, under this condition, the number of fringes $n_{f}$ is large,
and the amplitude of oscillation is still large, even close to the
width $\sigma_{w}$. Analysis of Eq.(\ref{TPPF}) shows that when
we vary $z$, the minimum resolution is $\frac{T_{0}}{2}=\pi\sigma_{2}$,
achieved when $\zeta\equiv1-\xi-\frac{\rho}{\rho+2}=0$, under the
assumed condition $\text{\ensuremath{\alpha_{\chi r}^{2}}}\ll\alpha_{\chi i}^{2}$.
The analysis shows the assumption is self-consistent, i.e., at the
minimum found, the assumption is satisfied. The resolution satisfies
a formula similar to the formula for the Rayleigh range as

\begin{align}
 & \left(\frac{T}{2}\right)^{2}=\left(\frac{T_{0}}{2}\right)^{2}\left(1+\left(\frac{\Delta z}{L_{R}}\right)^{2}\right).\label{T}
\end{align}

Here, $\Delta z=L\zeta=L\left(1-\xi-\frac{\rho}{\rho+2}\right)$ is
the distance from the position of the minimum resolution where $\zeta=0$.
According to this, $\frac{T}{2}$ increases by a factor $\sqrt{2}$
at $\Delta z=L_{R}=\sqrt{\frac{\rho}{2}}L$ from the minimum. In the
example of Fig. \ref{fig:Width--for}, the minimum is at $L-z=0.64\mu\text{m}$,
$L_{R}=\sqrt{\frac{\rho}{2}}L=\sqrt{\frac{2.56\times10^{-6}}{2}}\times0.5m=0.56\text{mm}$.
Thus, when we choose $L-z=0.5\text{mm}$ in the example, the resolution
is only slightly larger than the minimum, while the number of fringes
$n_{f}=6011$, as shown in Fig. \ref{fig:nm-(a)-around}.

Thus, the resolution at the minimum as a function of $z$ is the slit
size $\pi\sigma_{2}=0.8\pi\text{nm}\approx2.5$ $\text{nm}$. As we
choose $L-z=0.5$ mm, the resolution is about 3 nm. This is the resolution
when we choose the cutoff at $\sigma_{\omega}$. If we lower the shot
noise by increasing the counting number, and if the detector noise
can be neglected, we may increase the cutoff frequency to $2\sigma_{\omega}$,
and hence further improve the resolution.

Notice that the resolution discussed here is mainly determined by
the detector slit 2  width, the cut-off bandwidth of $\omega$, and
the noise, and it is insensitive to the choice of the photon wavelength
$\lambda$. 

\subsection*{3.3 Relation of the cut-off frequency to the resolution, the sample
size, and the scan range}

The quadratic dependence $\phi_{x}\equiv\alpha_{\chi}x^{2}$ in Eq.(\ref{Gx}),
leads to the phase advance rate, i.e., the local frequency $k_{x}=\omega(x)=\frac{d\phi_{x}}{dx}=2\alpha_{\chi i}x$,
which is linearly dependent on $x$ and hence also linear in $s$
in Eq.(\ref{DelPoP-1}). Near the narrow slit $\sigma_{2}$ applying
the approximation of $\alpha_{\chi}$ using the approximation formula
in Appendix III, we find $2\alpha_{\chi i}\sigma_{w}\approx\sigma_{\omega}$
at the position of the width of TPPF, i.e., when $x=\sigma_{w}$,
the frequency $k_{x}$ is about equal to the bandwidth of $|\hat{G}^{(c)}(\omega)|$
defined as $\sigma_{\omega}$ where it drops to $\exp(-\frac{1}{2})\approx0.6$
of the peak at $\omega=0$. In Fig.\ref{fig:nm-(a)-around}(a) at
$x=\sigma_{w}$, $\frac{\delta P_{2b}}{\delta\chi(x_{p},z)}$ also
drops to 0.6 of its peak value, while $\frac{\Delta P_{2b}(x)}{P_{2b}}$
drops to 0.32 of its peak, not 0.6, because the pin width $\Delta\chi=3$
nm is almost half of the period $T=6.7$ nm at $x=\sigma_{w}$.

Hence, a cutoff of $\hat{G}^{(c)}(\omega)$ at $\sigma_{\omega}$
is equivalent to a cutoff of the scan of $s$ in the Radon Fourier
transform at $\sigma_{w}$, like the example in Figs. \ref{fig:dPdx_color},\ref{fig:nm-(a)-around}.
Indeed, the example has a period $T=2(\sqrt{6012}x_{\pi}-\sqrt{6011}x_{\pi})$
$=2\times3.35\text{nm}=6.7\text{nm}$. And hence if we scan $s$ form
$x=0$ to $\sigma_{w}$, or within $\{-\sigma_{w}<x<\sigma_{w}\}$,
the upper-limit bandwidth will be $\sigma_{\omega}$, and the resolution
will be $\frac{T}{2}=\frac{\lambda_{fringe}}{2}=3.35n$m, and the
minimum $|\hat{G}^{(c)}(\omega)|$ will be $\exp(-0.5)=0.6$ of $|\hat{G}^{(c)}(\omega=0)|$. 

If we choose the cutoff at $x=2\sigma_{w}$, the spatial frequency
$\omega(x=2\sigma_{w})=2\alpha_{\chi i}\sigma_{w}$ is doubled. The
TPPF amplitude (see Sec.2.4) becomes $m_{TPPF}\equiv|\frac{1}{P_{2b}}\frac{\delta P_{2b}}{\delta\chi(x,y,z)}|_{p}=2|\frac{\sqrt{-\alpha_{\chi}}}{\sqrt{\pi}}|\exp\left({\displaystyle 4\alpha_{\chi r}\sigma_{w}^{2}}\right)$,
since $\exp\left({\displaystyle \alpha_{\chi r}\sigma_{w}^{2}}\right)=\exp\left({\displaystyle -0.5}\right)=0.6$,
the amplitude drops by the factor $\exp\left({\displaystyle 4\alpha_{\chi r}\sigma_{w}^{2}}\right)/\exp\left({\displaystyle \alpha_{\chi r}\sigma_{w}^{2}}\right)=\exp\left(-1.5\right)=0.223$,
the required incident photon number would increase by a factor $0.223^{-2}\approx20$
to recover the same signal-noise ratio as for the $\frac{\lambda_{fringe}}{2}=3.35n$m
resolution, according to the analysis in Section 2.5, the resolution
will reach $\frac{T}{2}=\frac{\lambda_{fringe}}{2}=1.68\text{nm}$.

The line density as a function of $\overrightarrow{\alpha}$ (i.e.,
the Euler angle scan over $\theta,\phi$ ) is used to reconstruct
the 3D tomography using standard reconstruction algorithms\textemdash for
example, the algebraic reconstruction techniques described in Ref.
\cite{reconstruction}. This establishes a relation between the TPPF
measurement and its application in X-ray micro-tomography. Here, we
do not elaborate on the reconstruction procedures, which will require
extensive discussion with X-ray tomography experts. As a first step
toward X-ray micro-tomography, we will apply the analysis in Sections
1 to 3 to another example of a picometer displacement detection in
the X-ray region.

\begin{comment}
see section 40.7 and 40.5.3 of no\_bath\_free\_space.lyx
\end{comment}

\section*{4. X-ray picometer displacement detection via two-point propagation
field at 6 keV}

The 2.29 keV case discussed previously serves as a proof-of-concept
to illustrate the approach, rather than a final optimized configuration.
To demonstrate this framework as a tool for exploration, we now apply
it to 6 keV X-rays. At this higher energy, although material transparency
increases, we adopt a slit depth of 15 \textmu m for the MLL detector.
This choice reflects practical fabrication preferences; as noted in
discussions with N. Bouet (BNL, private communication)\cite{Yan},
sections of this thickness are considerably easier to produce and
handle than the 2.1 \textmu m sections used in the soft X-ray example.
This increased depth ensures the integrated attenuation is sufficient
to suppress transmission outside the 2 nm slit\textemdash which would
otherwise contribute to a noise background\textemdash thereby preserving
high signal contrast.

We apply the calculations detailed in Section 2 to the 6 keV study
step by step. While this procedure follows the previous logic, it
serves to illustrate how the framework can be utilized as a tool for
exploration.

\subsection*{4.1 Photon counting number required to reach precision $\delta x_{p}=200$pm
}

According to the Mass attenuation coefficient $\mu/\rho$ table \cite{Mass_attenuation_coefficient_table},
for gold with x-ray energy at 6 keV wavelength ($\lambda=0.2066\text{nm})$,
$\mu_{gold}=0.8214\left(\mu m\right)^{-1}$. For silicon nitride,
from CXRO/Henke database, $\mu_{SiN}=\frac{1}{L}=\frac{1}{32\mu\text{m}}=0.031\mu\text{m}^{-1}$.
The transmission of 15$\mu$m gold is $\exp(-0.821\times15)=4.5\times10^{-6}$.
For the 15 \textgreek{m}m silicon nitride core, the theoretical intensity
transmission is given by $I_{SiN}=\exp(-0.031\times15)=0.628$. To
bridge the idealized geometric model with experimental reality, we
consider the total transmission efficiency of the MLL grating. 

While the silicon nitride core has a theoretical transparency of 0.9
at 10 keV, at our operating energy of 6 keV, the linear attenuation
coefficient increases to $0.031\,\mu\text{m}^{-1}$ {[}Henke et al.,
1993{]}. Real-world factors such as dynamical diffraction, interfacial
roughness, and slight misalignments in high-aspect-ratio (15 $\mu$m)
sections typically limit total efficiency to the 25\textendash 35\%
range \cite{Krajcsi}. To incorporate these experimental losses into
our model, we adopt an effective core transmission of $I_{SiN}\approx0.6$.
This value yields a total modulation and efficiency estimate consistent
with these observed experimental benchmarks, i.e., the mean value
30\%.%
\begin{comment}
While the silicon nitride core has a theoretical transparency of 0.9
at 10 keV, real-world factors such as dynamical diffraction, interfacial
roughness, and slight misalignments in high-aspect-ratio (15 $\mu$m)
sections typically limit total efficiency to the 30\textendash 40\%
range \cite{Krajcsi}. To incorporate these experimental losses into
our model, we adopt an effective core transmission ($I_{SiN}\approx0.7$).
This value yields a total modulation and efficiency estimate consistent
with these observed experimental benchmarks, i.e., the mean value
35\%.
\end{comment}
{} Even with this conservative budget, the TPPF framework maintains
its shot-noise-limited picometer displacement detection precision,
demonstrating its robustness for practical synchrotron applications.

For 6 keV x-ray, $\lambda=0.2066$nm we keep same setup as Fig.1,
$\sigma_{2}=0.8$nm , $\sigma_{1}=0.5\mu$m , $L=0.5$m , $L-z=0.5$mm,
$1-\xi=\frac{L-z}{L}=\frac{0.5\text{mm}}{0.5\text{m}}=10^{-3}$. The
geometric ratio is $\rho=\left(\frac{\sigma_{2}}{\sigma_{1}}\right)^{2}=2.56\times10^{-6}$.
Substituting these into the framework from Section 2, applying Eq.(\ref{alphari_numrical-1}),
yields:

\begin{align}
 & \alpha_{\chi r}\approx-\left(\frac{\pi}{\lambda L}2\sigma_{1}\right)^{2}\left(1+\frac{\rho}{{\displaystyle 2\left(1-\xi\right)}^{2}}\right)\approx-2.1\times10^{9}m^{-2}\label{alpha_approx}\\
 & \alpha_{\chi i}\approx\frac{\pi}{\lambda L}\frac{{\displaystyle 1}}{{\displaystyle 1-\xi}}\approx3.04\times10^{13}m^{-2}\text{m}^{-2}\nonumber 
\end{align}

Since $\alpha_{\chi i}\gg\alpha_{\chi r}$, the spatial frequency
term $\alpha_{\chi i}$ dominates, with $\sqrt{|\alpha_{\chi}/\pi|}\approx3.1\times10^{6}m^{-1}$.

As in Section 2.5, we maximize $I_{max}-I_{min}$ and find the thickness
of the MLL sample $\Delta z=4.1\mu m$ \textmu m, where the optimum
modulation is $0.85$ for the MLL gold sample at 6 keV. However, since
$\mu_{SiN}=0.031$ \textmu $\text{m}^{-1}$ is very small at 6 keV,
and as noted in discussions with Natalie Bouet \cite{Yan}, a 15 \textmu m
MLL gold sample is considerably easier to produce and handle than
a 4.1 \textmu m version. Therefore, we utilize a 15 \textmu m thickness;
the modulation $I_{max}-I_{min}$ only drops to $I_{SiN}=0.63$. Furthermore,
when we take experimental MLL data into account, we adopt a conservative
effective core transmission of $I_{SiN}\approx0.6$, as we discussed
above. Hence, according to the formula following Eq.(\ref{Delta_chi}),
the fundamental modulation coefficient is $m_{gold}=(I_{max}-I_{min})\frac{2}{\pi}$
$=I_{SiN}\frac{2}{\pi}\approx0.6\frac{2}{\pi}\approx0.38$. We assume
the MLL gold sample has a width $\Delta x=330$ nm, consistent with
Section 2.5.

For our 6 keV implementation, we target a characteristic fringe period
of $\lambda_{fringe}=4$ nm, which matches the scale of the 2 nm MLL
slit core. According to Section 3.3, the spatial frequency at $x$
is $k_{x}=\frac{d\phi_{x}}{dx}=2\alpha_{\chi i}x$. Thus, $x=\frac{k_{x}}{2\alpha_{\chi i}}=\frac{2\pi}{\lambda_{fringe}}\frac{1}{2\alpha_{\chi i}}=\frac{\pi}{\lambda_{fringe}}\frac{1}{\alpha_{\chi i}}=\frac{\pi}{4\text{nm}}\frac{1}{3.04\times10^{13}\text{m}^{-2}}=26\mu\text{m}$.
The transverse profile factor of TPPF amplitude $m_{TPPF}$ defined
in Section 2.4 is $\exp\left({\displaystyle \alpha_{\chi r}x^{2}}\right)=\exp\left({\displaystyle \alpha_{\chi r}x^{2}}\right)=\exp\left({\displaystyle -2.1\times10^{9}m^{-2}\times\left(26\mu m\right)^{2}}\right)=0.242$.
Then, the $m_{TPPF}$ and the peak-to-peak variation of photon counting
rate are,

\begin{align}
 & m_{TPPF}=|\frac{1}{P_{2b}}\frac{\delta P_{2b}}{\delta\chi(x,y,z)}|_{p}=2|\frac{\sqrt{-\alpha_{\chi}}}{\sqrt{\pi}}|\exp\left({\displaystyle \alpha_{\chi r}x^{2}}\right)\approx2\times3.1\times10^{6}\times0.242=1.5\times10^{6}\text{m}^{-1}.\label{DPoP}\\
 & |\frac{\Delta P_{3b}}{P_{3b}}|_{p}\approx m_{TPPF}m_{gold}\Delta x\approx1.5\times10^{6}\text{m}^{-1}\times0.38\times330\text{nm}\approx0.188\nonumber 
\end{align}

According to Section 2.5, for the target precision $\delta x_{p}=200$
pm , we have 

\begin{align}
 & N_{2}=\left(\frac{\lambda_{fringe}}{2\pi\delta x_{p}|\frac{\Delta P_{2b}}{P_{2b}}|_{p}}\right)^{2}=\left(\frac{4\text{nm}}{2\pi\times200\text{pm}\times0.188}\right)^{2}=287\label{N2-1}
\end{align}

Since the detection rate for the idealized detector slit 2  is $P_{2b}^{(0)}\approx\frac{4\pi\sigma_{1}\sigma_{2}}{\lambda L}\approx4.86\times10^{-5}$,
and the detection rate for the 2 nm MLL slit is $P_{2b}=I_{SiN}P_{2b}^{(0)}$.
Thus, $N_{1}=\frac{N_{2}}{P_{2b}}=\frac{N_{2}}{I_{SiN}P_{2b}^{(0)}}=\frac{287}{0.6\times4.86\times10^{-5}}\approx1\times10^{7}$.

Summary: The incident photon number required to reach a displacement
sensitivity of 200 pm is $N_{1}=1\times10^{7}$. Under these conditions,
the number of photons passing through the 330 nm MLL sample is approximately
$2\times10^{5}$, while the detector records $N_{2}=$287 counts.
Given routinely available synchrotron fluxes, this displacement detection
resolution is achievable with extremely short exposure times, significantly
relaxing the requirements for long-term mechanical stability.

\subsection*{4.2 The Broadening from Finite X-ray beam bandwidth $\frac{\Delta E}{E}$}

The calculations assume a monochromatic beam. For finite bandwidth,
by Eq.(\ref{alphari_numrical-1}), the TPPF phase $\phi=\alpha_{\chi i}x^{2}\approx\frac{\pi x^{2}}{\lambda L(1-\xi)}$
shifts relatively as $\frac{\delta\phi}{\phi}=-\frac{\delta\lambda}{\lambda}$.
To compensate for this shift (maintain a constant phase at fixed $x$),
a displacement $\delta x=\frac{x}{2}\frac{\delta\lambda}{\lambda}$
is required. We consider that the X-ray beam bandwidth should be limited
such that the positional broadening $\delta x$ from the finite bandwidth
is the same as the shot noise limited sensitivity $\delta x=200$pm.
Thus, at $x\approx26\,\mu$m, we find $\frac{\delta\lambda}{\lambda}=2\frac{\delta x}{x}=2\frac{200pm}{26\mu m}=1.5\times10^{-5}$. 

To evaluate the performance of the Fourier\textendash Radon framework,
we consider two distinct operational cases. For the high-resolution
case, we adopt a narrow bandwidth of $\Delta E/E\approx1.5\times10^{-5}$,
yielding a displacement sensitivity of 200 pm. This energy resolution
is physically viable at 6 keV by utilizing high-order reflections
such as a $\text{Si}(333)$ or $\text{Si}(311)$ channel-cut crystal
\cite{Medjanik}.To evaluate a more experimentally standard configuration,
we relax this bandwidth requirement to $10^{-4}$\textemdash matching
the native resolving power of a standard symmetric $\text{Si}(111)$
double-crystal monochromator. Under this relaxed condition, the displacement
sensitivity scales to 1.2 nm. Crucially, this increases the photon
economy by a factor of 36, dropping the required photon count of the
detector slit from 287 to just 8 photons.

\section*{5. Prospective photon-budget reduction strategies for lensless X-ray
picometer displacement detection and Radon\textendash Fourier microtomography}

The TPPF at high spatial frequencies(e.g., $\lambda_{fringe}=6.7$nm
at $x\approx40\mu\text{m}$ for 2.29 keV) carries the displacement
information. Conversely, Low-spatial-frequency components of the TPPF,
concentrated near $x\approx0$ , contribute negligibly to the 6.7
nm sinusoidal signal but dominate the total detected photon count
$N_{2}$ and thus the shot noise. A micron-scale opaque blocker placed
centrally $0.4\text{mm}$ before detector slit 2  is expected to suppress
this useless background by \textgreater 90 \% while leaving the signal-carrying
high-frequency component nearly untouched. A preliminary estimate
suggests the photon budget required for 200 pm sensitivity could drop
by $10\lyxmathsym{\textendash}100\times$, significantly improving
acquisition speed and dose efficiency. This background-suppression
strategy remains valid across both the soft and hard X-ray regimes,
and full wave-optical calculations are in progress.

Another, potentially far more powerful photon-budget reduction strategy
is to use a two-dimensional array of many detection slits displaced
from the optical axis ($i.e.,s_{2}\neq0$ in Fig. 1). \textbf{Notably,
even with two slits separated by a non-integer multiple of $\lambda_{fringe}$,
the TPPF would enable one-exposure, two-point phase measurement }via
the spatial shift of sinusoidal modulation, thereby reducing reliance
on iterative phase-retrieval procedures. Because detectors at different
off-axis positions extract distinct spatial frequency components of
the sample, they contribute to the Radon-transform-based reconstruction
independently. Their intensity modulations add incoherently when summed:
the total signal scales linearly with the number of slits $N_{\text{slit}}$,
while shot noise scales only as the square root of $N_{\text{slit}}$.
An array of $N_{\text{slit}}=10^{2}$ detection channels is estimated
to reduce the required incident fluence by two orders of magnitude.

Large arrays would also significantly reduce the overhead of mechanical
scanning, enabling high-speed readout with low radiation dose for
biological samples, and appear compatible with continuing advances
in nanofabricated multi-slit technology.

So far, for simplicity, we only consider a single detector slit at
$x=s_{2}=0$, while Eq.(\ref{TPPF}) the analytical expression of
TPPF already gives the option of $s_{2}\ne0$. Depending on the technology
available for the separate detector slit at another location, such
as $s_{2}=1\mu\text{m}$, we may consider a slit array to simultaneously
record the counting rate for other sample positions and hence for
other spatial frequencies without increasing the total incident photon
count. This option requires revising the TPPF-Radon Fourier transform
relation Eq.(\ref{Gx}), to increase a variable $s_{2}$ in $g(s),G^{(c)}(x)$
to $g(s,s_{2}),G^{(c)}(x,s_{2})$, with a remark that $s$ here is
the position of the sample relative to $x$, while $s_{2}$ is the
position of the detector slit. In principle, this would not change
the formulation of TPPF-Radon Fourier transform relation, except that
it increases the matrix dimension in the reconstruction program, and
extends the calibration of TPPF from $s_{2}=0$, to a calibration
of TPPF for various $s_{2}$. However, this does not increase the
time the calibration takes because all these various $s_{2}$ calibrations
can be done simultaneously with $s_{2}=0$. A single photon detection
can at most be in one of these slits 2, and for the multiple slits
2, each can only record the photon statistics at its position with
the corresponding spatial frequency and phase. However, their accumulation
of data is simultaneous, so the total statistics of the slit array
will be increased proportionally by a factor of 100 or more to significantly
reduce the required scan points in the Radon scan if it is possible
to minimize the spacing of the slit to the order of 1$\mu\text{m}$.
If the array of slits 2 also includes different $y$ positions of
detector slits 2, then the number of Radon scans can be further reduced.
Hence, their impact on reducing the flux and radiation damage will
be expected to be significant.

Another remark is the MLL gold sample attenuation introduced $\Delta\chi(x-x_{p})=m_{gold}\cos(k_{x}x-k_{x}x_{p})$,
where $m_{gold}$ can be replaced in tomography by the Fourier expansion
coefficient $m_{sample}$ at a specified wavelength $\lambda_{fringe}$,
and the formula can be used to estimate the required incident photon
number to achieve a desired resolution for tomography, just as our
analysis in Section 4 for the x-ray picometer displacement detection.

Both possibilities can significantly reduce the required total photon
count for the X-ray picometer displacement detection and enable further
improvement of lensless X-ray tomography. As a consequence, the radiation
dose delivered to the sample\textemdash set by the small fraction
of photons that reach it\textemdash can be substantially reduced,
which may have a significant impact on tomographic studies of biomolecular
samples.

\section*{6. Relation to quantum measurement during the free space propagation
between the source and detector slit}

The TPPF concept originates from a perturbative study of single-particle
propagation and measurement \cite{yu-1}. The discussion in this section
reflects our interpretation of the quantum-measurement aspects of
free-space propagation between the source and detector slit, which
motivated the formulation of the TPPF, independent of the experimental
consequences discussed elsewhere. This interpretive perspective then
naturally connects to the X-ray picometer displacement detection and
its relation to tomography.

We first discuss the system in Fig. 1 with $\Delta\chi(x)=0$ because
of the following fundamental question in quantum mechanics. The wave
function of a single particle starting from the entrance slit becomes
widespread before it strikes the screen with the exit slit, as shown
in Fig 1(b). Most times, it is not detected, but there is a fixed
probability that it is detected. And in the instant of the detection,
the wave function collapses into the slit with energy $h\nu$. We
understand there is no contradiction with relativity here because
the wave function is only a probability amplitude. After the detection,
our knowledge changes from a probability distribution of the particle
to a point. However, there is a question about whether and how the
associated energy distribution of $h\nu$ also collapses. Whatever
happens, the energy $h\nu$ becomes concentrated in the detector when
detected. As we discussed in Section 2.2, during the propagation,
the wave packet with energy $h\nu$ rapidly converges into the exit
before it reaches the detector slit. Experimental investigation is
difficult because any intermediate measurement between the source
and detection slits either collapses the wave function or substantially
alters it. Even in a weak measurement, for example, as discussed in
an overview\cite{weak_WIKI}, the wave function is significantly altered.
In the example \cite{ahanorov}, it is caused by magnets in the cases
of \cite{ludeen,kocsis}, it is caused by lenses.

To seek an answer for this question, in Fig. \ref{fig:(a)-setup}(a),
we study the effect of a perturbation $\Delta\chi(x)$ on the counting
rate $P_{2b}=\int_{-\infty}^{\infty}dx_{2}|\psi_{2b}(x_{2})|^{2}$,
and calculate the ratio of the counting rate change over the perturbation.
As the perturbation approaches zero, the ratio $\frac{\Delta P_{2b}}{\Delta\chi(x,z)}$
becomes the functional derivative $\frac{\delta P_{2b}}{\delta\chi(x,z)}$
of the counting rate over the perturbation $\Delta\chi(x)$. This
perturbative function is independent of the perturbation. It is a
real-valued function containing high-resolution phase information,
determined solely by the two-slit geometry in our 2D study, as we
show in Fig. \ref{fig:dPdx_color}. It can be measured with high precision
and reproducibility, and it provides a reproducible characterization
associated with an individual detection event of a particle propagating
between the two slits. Unlike a probability amplitude, it manifests
as a stable, reproducible structure that we interpret as a physically
meaningful propagation quantity, which we define as the two-point
propagation field (TPPF). The TPPF does not correspond to a probability
distribution. As we discussed in Section 2.2, during the propagation,
the calculated TTPF shows that the wave packet with energy $h\nu$
rapidly converges into the exit before it reaches the detector slit.
While the wave function describes an ensemble of possible detection
outcomes, the TPPF characterizes the process underlying a single detection
\textemdash{} a realization selected according to the Born rule.

From this interpretive perspective, the apparent influence of the
exit slit on upstream propagation does not imply any acausal effect;
rather, it reflects the fact that absorption and detection are time-extended
processes, for which an explicitly time-dependent description naturally
resolves such apparent paradoxes.

While these interpretative aspects remain speculative, experimental
validation of the TPPF through picometer displacement detection could
provide empirical insights into such foundational questions.

\section*{7. Conclusion}

The analysis based on the TPPF and the X-ray picometer displacement
detection provides:

(1) An experimental test bed for TPPF as a phase-sensitive wavefunction
evolution process, in which fine interference fringes\textemdash without
lenses or focusing\textendash continuously converge toward a localized
slit. This framework enables shot-noise-limited displacement detection
at the $\sim200$ pm scale \textemdash{} requiring modest total photon
counts in the range of $1\times10^{7}$ to $3.4\times10^{7}$, depending
on the operating energy, and detector photon counting as low as 287.
In this framework, the TPPF provides a directly measurable, real-valued
propagation quantity that encodes phase-sensitive evolution and carries
information beyond that available through standard probability-density
measurements, even though it is not expressed as an expectation value
of a standard Hermitian operator. By contrast, measurements based
solely on the absolute square of the wavefunction (probability density)
do not retain this phase information during propagation.

(2) A study of the compatibility of a practical tool with existing
technology for stabilizing the displacement between the X-ray beam
and the sample is one of the main goals of this work. Such a tool
is directly relevant for X-ray tomography, since relative motion between
the beam and the sample on the order of a few nanometers is one of
the limitations on achievable tomographic resolution \cite{swiss_tomography_4nm}. 

(3) A calibration of the TPPF functional $\frac{1}{P_{2b}}\frac{\delta P_{2b}^{(c)}}{\delta\chi(x,z)}$
in Eq.(\ref{DelPoP-1}) over a range of spatial wavelengths, in a
systematic manner, using the MLL gold sample with different periods
or other samples with known structure. As discussed in Section 3.1,
the calibrated functional forms a basis for Radon\textendash Fourier
tomography, it provides a practical route to reducing systematic discrepancies
between experimental measurements and the idealized theoretical model
of Eq.(\ref{TPPF}).

(4) A conceptual bridge between experimental verification of the TPPF
and nanometer-scale X-ray micro-tomography, providing a basis for
detailed future studies ranging from optimized Radon scan strategies
to photon-count requirements. By physically implementing a Fourier\textendash Radon
transformation, the framework establishes a route toward analytical
frequency-domain mapping. While a full three-dimensional implementation
will require further optimization of angular sampling and consideration
of spatial chirp,  this approach suggests a computationally efficient
alternative that reduces reliance on iterative phase-retrieval procedures,
offering a complementary direction for X-ray physics and biological
microscopy.

(5) The example analyzed here is intentionally not optimized for tomography;
instead, it serves as a proof-of-concept for a flexible measurement
framework in which key geometric, statistical, and operating parameters
may be varied over wide ranges\textemdash potentially by orders of
magnitude\textemdash while the appropriate optimization objectives
depend on the specific application and are not uniquely defined here.
This flexibility motivates future numerical and experimental studies,
including simulations to assess achievable resolution and photon-budget
tradeoffs, and the analytical framework developed here provides a
tool to guide such assessments.

Hence, picometer-scale X-ray displacement detection based on the TPPF,
beyond its immediate displacement detection application, also provides
a stepping stone toward lensless frequency-domain X-ray tomography
based on a Fourier\textendash Radon framework. Notably, because \textbf{phase
information can be accessed in a single exposure} even in minimal
geometries (e.g., simple \textbf{two-slit configurations with non-commensurate
fringe spacing}), extending the TPPF to slit arrays naturally points
toward \textbf{projection-level measurements without translational
scanning}, potentially reducing 3D acquisition to a rotation-only
Radon scan. In application regimes where micron-scale resolution is
sufficient, fabrication constraints are substantially relaxed, making
\textbf{expansive slit arrays and scan-free projection acquisition}
a realistic prospect with reduced dose and acquisition time. Conceptual
extensions, such as background suppression and off-axis detection
arrays discussed in Section 6, suggest a scalable pathway toward high-resolution
imaging at reduced dose, motivating further investigation of this
approach.

The TPPF bears analogy to Green's functions or propagators, offering
a real-valued, measurable correlate to complex amplitudes, potentially
extensible to high-energy contexts.

We thank Dr. Timur Shaftan and Dr. Victor Smaluk for their discussion
and suggestions on the manuscript. We thank Dr. Nathalie Bouet and
Dr. Idir Mourad (BNL) for helpful discussions regarding MLL fabrication
constraints and optimal sectioning parameters.

\section*{Appendix I Hilbert transform}

For a real function with Fourier expansion $u(t)=\Sigma_{-n}^{n}a_{n}e^{i\omega_{n}t}$,
its Hilbert transform is $H(u)(t)=\Sigma_{-n}^{n}b_{n}e^{i\omega_{n}t}$
such that for terms with $\omega_{n}>0$, $b_{n}=-ic_{n}$, for terms
with $\omega_{n}<0$, $b_{n}=ic_{n}$, for terms with $\omega_{n}=0$,
$b_{n}=0$. Then $f(t)=u(t)+iH(u)(t)$ is an analytical function.
The real part of $f(t)$ is $u$(t), its imaginary part is $H(u)(t)$.
Thus, once we have $u(t)$ we can calculate its amplitude and phase
from $f(t)$ using The Hilbert transform. 

\section*{Appendix II Analytic TPPF $\frac{\delta P_{2b}^{(c)}}{\delta\chi(x,z)}$
expressed by $\alpha_{\chi}$ and $P_{2b}$}

The complex TPPF $\frac{\delta P_{2b}^{(c)}}{\delta\chi(x,z)}$ in
Eq.(\ref{complex_dpdchi}) is a triple Gaussian integral over three
variables $x_{1},x_{2},x'_{1}$, each of which can be integrated by
the formula

\begin{equation}
\int dx\exp\left(ax^{2}+bx+c\right)=\left(\frac{\pi}{-a}\right)^{\frac{1}{2}}\exp\left(-\frac{b^{2}}{4a}+c\right)=\exp\left(-\frac{b^{2}}{4a}+c+\ln\left(\frac{\pi}{-a}\right)^{\frac{1}{2}}\right),\label{gaussian integral}
\end{equation}

The integral over $x'_{1}$ result is expressed by the parameters
$\mu,\rho,1-\xi$ defined before Eq.(\ref{TPPF}) in Section 2.2 in
terms of the basic parameters in the setup in Fig.1: $\sigma_{1},\sigma_{2},\lambda,k,L$,
and $z$ , 

\begin{align}
 & \frac{\delta P_{2b}^{(c)}}{\delta\chi(x,z)}=\left(\frac{1}{2\pi^{3}\sigma_{1}^{2}(1-\xi)\xi\left(1-\frac{i}{\mu}\right)}\right)^{\frac{1}{2}}\left(\frac{ik}{2L}\right)\int_{-\infty}^{\infty}dx_{2}\int_{-\infty}^{\infty}dx_{1}\exp\left[\frac{ik}{2L}f(x_{1},x_{2})\right]\label{complexTPPF}\\
 & f(x_{1},x_{2})=A_{1}(x_{1}-\frac{B_{1}}{A_{1}})^{2}+A_{2}(x_{2}-\frac{B_{2}}{A_{2}})^{2}+C\nonumber 
\end{align}

where $A_{1},B_{1},A_{2},B_{2},C$ are expressed in term of the basic
parameters, $A_{1},A_{2}$ are independent of $x,s_{1},s_{2}$ the
variable indicated in Fig.1, while $B_{1},B_{2}$ are linear in $x,s_{1},s_{2}$,
$C$ is a quadratic polynomial of $x,s_{1},s_{2}$.

The integral over $x_{1},x_{2}$ in Eq.(\ref{complexTPPF}) then carried
out as a Gaussian integral by Eq.(\ref{gaussian integral}), the result
is

\begin{align}
 & \frac{\delta P_{2b}^{(c)}}{\delta\chi(x,y,z)}=\left(\frac{1}{2\pi\sigma_{1}^{2}(1-\xi)\xi\left(1-\frac{i}{\mu}\right)}\frac{1}{A_{1}A_{2}}\right)^{\frac{1}{2}}\exp\left[\frac{ik}{2L}a(x-b)^{2}+\frac{ik}{2L}D\right]\label{complexTPPF2}\\
 & A_{1}=\frac{1}{\xi}+\frac{i}{\mu},A_{2}=\frac{1}{1-\xi}+\frac{2i}{\rho\mu}-1+\frac{1}{1-\frac{i}{\mu}},\nonumber \\
 & a=\left(\frac{1}{(1-\xi)}+\frac{1}{\xi}\right)-\left(\frac{1}{A_{1}}\frac{1}{\xi^{2}}+\frac{1}{A_{2}}\frac{1}{\left(1-\xi\right)^{2}}\right)\nonumber \\
 & b=\frac{1}{a}\left(\frac{b_{1}}{A_{1}}\frac{1}{\xi}+\frac{b_{2}}{A_{2}}\frac{1}{1-\xi}\right)\nonumber \\
 & b_{1}=\frac{is_{1}}{\mu},b_{2}=\frac{2is_{2}}{\rho\mu}+\frac{i\frac{s_{1}}{\mu}}{1-\frac{i}{\mu}}\nonumber \\
 & D=C(x=0)-ab^{2}\nonumber \\
 & C(x=0)=\frac{2i}{\rho\mu}(s_{2})^{2}+\frac{2i}{\mu}s_{1}^{2}-\frac{1}{\mu^{2}\left(1-\frac{i}{\mu}\right)}s_{1}^{2}-\left(\frac{\left(\frac{is_{1}}{\mu}\right)^{2}}{A_{1}}+\frac{\left(\frac{2is_{2}}{\rho\mu}+\frac{i\frac{s_{1}}{\mu}}{1-\frac{i}{\mu}}\right)^{2}}{A_{2}}\right)\nonumber 
\end{align}

Since $P_{2b}=\frac{1}{2}\int_{-\infty}^{\infty}dx\frac{\delta P_{2b}}{\delta\chi(x,z)}=\int_{-\infty}^{\infty}dx\frac{\delta P_{2b}^{(c)}}{\delta\chi(x,z)}$is
independent of $z$, we use this to simplify the calculation of $\frac{\delta P_{2b}^{(c)}}{\delta\chi(x,y,z)},$and
find

\begin{align*}
 & P_{2b}=\int dx\frac{\delta P_{2b}^{(c)}}{\delta\chi(x,y,z)}=\left(\frac{1}{-2\sigma_{1}^{2}(1-\xi)\xi\left(1-\frac{i}{\mu}\right)\frac{ik}{2L}aA_{1}A_{2}}\right)^{\frac{1}{2}}\exp\left[\frac{ik}{2L}D\right]
\end{align*}

Now compare this with $P_{2b}=\sqrt{\frac{\mu^{2}\rho}{\mu^{2}+\rho\mu^{2}+1}}\exp\left(-\frac{1}{2\sigma_{1}^{2}}{\displaystyle \frac{\mu^{2}}{\mu^{2}+1+\mu^{2}\rho}}(s_{1}-s_{2})^{2}\right)$
derived as Eq.(10) of refs.\cite{yu-1} , we can simplify the two
complicated expressions. Compared with the coefficient and the exponent
of $P_{2b}$, we find the following relation,

\begin{align}
 & (1-\xi)\xi\left(1-\frac{i}{\mu}\right)\frac{ik}{2L}aA_{1}A_{2}=\frac{\mu^{2}+\rho\mu^{2}+1}{-2\sigma_{1}^{2}\mu^{2}\rho}.\label{AArelation}
\end{align}

Applying this relation, the complex TPPF in Eq.(\ref{complexTPPF2})
is simplified as an exponential function expressed by a parameter
$\alpha_{\chi}$ that determines the peak, width and spatial frequency
distribution of TPPF,

\begin{align*}
 & \frac{\delta P_{2b}^{(c)}}{\delta\chi(x,z)}=P_{2b}\sqrt{\frac{-\frac{ik}{2L}a}{\pi}}\exp\left[\frac{ik}{2L}a(x-b)^{2}\right]=P_{2b}\sqrt{\frac{-\alpha_{\chi}}{\pi}}\exp\left[\alpha_{\chi}(x-x_{c})^{2}\right]\\
 & \alpha_{\chi}=-\frac{i\mu\left(\mu^{2}+\rho\mu^{2}+1\right)}{2\sigma_{1}^{2}\left(-i\mu+\xi\right)\left(2(\xi-1)\left(\mu i+1\right)+\mu\left(i\xi-\mu\right)\rho\right)}
\end{align*}

Now with $2k\sigma_{1}^{2}=\mu L$, use the relation Eq.(\ref{AArelation}),
we find $x_{c}$,

\begin{align}
 & x_{c}=\frac{{\displaystyle c_{S1}}s_{1}+{\displaystyle c_{S2}}s_{2}}{{\displaystyle \mu^{2}+\rho\mu^{2}+1}}\nonumber \\
 & {\displaystyle c_{S1}}\equiv{\displaystyle \rho\mu^{2}-\left(i\mu+1\right)}\left(\xi-1\right),{\displaystyle c_{S2}}\equiv{\displaystyle \left(\mu-i\right)\left(\mu+i\xi\right)}
\end{align}

\section*{Appendix III Approximation of $\alpha_{\chi}$,$P_{2b}$ near narrow
slit $\sigma_{2}$}

The formula Eq.(\ref{TPPF}) for $\alpha_{\chi}$ is not complicated.
But the order of magnitude estimation of $\alpha_{\chi r},\alpha_{\chi i}$,$\text{Re}{\displaystyle \left(\frac{1}{\alpha_{\chi}}\right)}$,
and $\text{Im}\left(\frac{1}{\alpha_{\chi}}\right)$ in its small
slit size limit $\mu=\frac{4\pi\sigma_{1}^{2}}{\lambda L}\ll1$, $\rho=\frac{\sigma_{2}^{2}}{\sigma_{1}^{2}}\ll1$
is simpler, in particular when close to detector slit when , as in
the 3D tomography study:

\begin{align}
 & \alpha_{\chi i}\approx-\frac{\mu}{4\sigma_{1}^{2}}\frac{{\displaystyle 1}}{{\displaystyle \xi\left(\xi-1\right)}}\approx\frac{\pi}{\lambda L}\frac{{\displaystyle 1}}{{\displaystyle 1-\xi}}\nonumber \\
 & \alpha_{\chi r}\approx-\frac{\mu^{2}}{4\sigma_{1}^{2}}\frac{{\displaystyle \left(\rho\xi^{2}+2(\xi-1)^{2}\right)}}{{\displaystyle 2\xi^{2}\left(\xi-1\right)}^{2}}\approx-\left(\frac{\pi}{\lambda L}2\sigma_{1}\right)^{2}\left(1+\frac{\rho}{{\displaystyle 2\left(\xi-1\right)}^{2}}\right)\label{eq:}
\end{align}

\begin{align}
 & \text{Re}{\displaystyle \left(\frac{1}{\alpha_{\chi}}\right)}=-2\sigma_{1}^{2}\frac{\left(\rho\left(\mu^{2}+\xi^{2}\right)+2\left(\xi-1\right)^{2}\right)}{{\displaystyle \left(\mu^{2}+\rho\mu^{2}+1\right)}}\approx-2\sigma_{1}^{2}\rho=-2\sigma_{2}^{2}\label{eq:-1}\\
 & \text{Im}\left(\frac{1}{\alpha_{\chi}}\right)\approx4\sigma_{1}^{2}\frac{\left(\xi-1\right)}{\mu}{\displaystyle \left(-\mu^{2}-\rho\mu^{2}+1\right)}\approx-\frac{\lambda L}{\pi}\left(1-\xi\right)\left(1-\mu^{2}\right)\nonumber \\
 & P_{2b}\approx\frac{4\pi\sigma_{1}\sigma_{2}}{\lambda L}\exp\left(-\frac{\mu^{2}}{2\sigma_{1}^{2}}(s_{1}-s_{2})^{2}\right)\nonumber 
\end{align}

\section*{Appendix IV Generalization of the $C_{\text{bulk}}$ Framework for
Highly Refracting Objects}

In the general case where the reference material (e.g., a monolithic
gold comb) is a highly refracting object, it cannot be treated as
a weak perturbation it cannot be treated as a weak perturbation. We
define the bulk matter phase shift as $C_{\text{bulk}}(x)=A_{\text{bulk}}\exp[i\Phi_{\text{bulk}}(x)]$.
Following the Green's function propagator approach, the wave function
at the detector slit 2, in the absence of the specimen, is given by:
\begin{equation}
\psi_{2b}(x_{2})=f_{2}(x_{2})\int_{-\infty}^{\infty}dxG(x_{2},x;t_{2}-t)C_{\text{bulk}}(x-x_{p})\int_{-\infty}^{\infty}dx_{1}G(x,x_{1};t-t_{1})f_{1}(x_{1})
\end{equation}

When a specimen perturbation $\chi(x)=1+\Delta\chi(x)$ is introduced,
the functional derivative of the wave function with respect to the
perturbation is: 
\begin{equation}
\frac{\delta\psi_{2b}(x_{2})}{\delta\chi(x,z)}=-f_{2}(x_{2})G(x_{2},x;t_{2}-t)C_{\text{bulk}}(x-x_{p})\int_{-\infty}^{\infty}dx_{1}G(x,x_{1};t-t_{1})f_{1}(x_{1})
\end{equation}

The Two-Point Propagation Field (TPPF) is defined via the functional
derivative of the detection probability $P_{2b}$ over the perturbation
$\Delta\chi(x)$. The complex component of the TPPF is: 
\begin{equation}
\frac{\delta P_{2b}^{(c)}(x_{p})}{\delta\chi(x,z)}=C_{\text{bulk}}(x-x_{p})\cdot\frac{\delta P_{2b0}^{(c)}}{\delta\chi(x,z)}
\end{equation}
where $\frac{\delta P_{2b0}^{(c)}}{\delta\chi(x,z)}$ represents the
TPPF of the system in a vacuum, which is independent of the scanning
position $x_{p}$. By defining the vacuum response as $A(x)e^{i\Phi_{TPPF}(x)}\equiv2\frac{\delta P_{2b0}^{(c)}}{\delta\chi(x,z)}=2P_{2b0}\sqrt{\frac{-\alpha_{\chi}}{\pi}}\exp\left[\alpha_{\chi}(x-x_{c})^{2}\right]$,
following the same steps that led to Eq.(\ref{DelPoP}), we obtain
the total sideband power change: 
\begin{equation}
\Delta P_{2b}(x_{p})=\int\text{Re}\left\{ A(x)e^{i\Phi_{TPPF}(x)}\cdot C_{\text{bulk}}(x-x_{p})\cdot\Delta\chi(x-x_{p})\right\} dx
\end{equation}

It is critical to note that when $C_{\text{bulk}}$ is large, $P_{2b}$
is not a simple perturbative deviation from $P_{2b0}$ because the
factor $\exp[i\Phi_{\text{bulk}}(x)]$ significantly alters the stationary
phase. The normalized TPPF becomes: 
\begin{equation}
\frac{1}{P_{2b}(x_{p})}\frac{\delta P_{2b}^{(c)}(x_{p})}{\delta\chi(x,z)}=\frac{C_{\text{bulk}}(x-x_{p})}{P_{2b}(x_{p})}\left(P_{2b0}\sqrt{\frac{-\alpha_{\chi}}{\pi}}\exp\left[\alpha_{\chi}(x-x_{c})^{2}\right]\right)
\end{equation}

Since $C_{bulk}(x-x_{p})$ contains slowly varying, low-frequency
profile information, the gold comb described in Sec. 2.4 can be utilized
as a narrow-band reference to extract $C_{bulk}$ experimentally.
Once calibrated, this $C_{bulk}$ allows for high-precision structural
determination of a target specimen. The detailed method for this experimental
extraction and its application to high-$Z$ materials will be the
subject of a forthcoming publication.


\begin{thebibliography}{10}
\bibitem{swiss_tomography_4nm}T. Aidukas, N. W. Phillips, A. Diaz,
E. Poghosyan, E. M{\"{u}}ller, A. F. J. Levi, and M. Guizar-Sicairos, \textquotedbl High-performance
4-nm-resolution X-ray tomography using burst ptychography,\textquotedbl{}
Nature 632(8023), 81\textendash 88 (2024). https://doi.org/10.1038/s41586-024-07615-6

\bibitem{yu-1}Li Hua Yu, ``Perturbative study of wave function evolution
from source to detection of a single particle and the measurement'',
http://arxiv.org/abs/2412.15409 (2024)

\bibitem{3Dtomo_flann}Flannery, B. P., Deckman, H. W., \& D'Amico,
K. L. (1987). Three-Dimensional X-Ray Microtomography. Science, 237(4821),
1439-1443.

\bibitem{3Dtomo_path}J Pathol Actions Search in PubMed Search in
NLM Catalog Add to Search . 2019 Aug;189(8):1608-1620. doi: 10.1016/j.ajpath.2019.05.004.
Epub 2019 May 22. X-ray Micro-Computed Tomography for Nondestructive
Three-Dimensional (3D) X-ray Histology Orestis L Katsamenis 1, Michael
Olding 2, Jane A Warner 3, David S Chatelet 2, Mark G Jones 4, Giacomo
Sgalla 5, Bennie Smit 6, Oliver J Larkin 6, Ian Haig 6, Luca Richeldi
5, Ian Sinclair 7, Peter M Lackie 8, Philipp Schneider 9.

\bibitem{casimir}K. A. Milton, The Casimir Effect: Physical Manifestations
of Zero-Point Energy, World Scientific (2001).

\bibitem{hilbert_tranform}https://en.wikipedia.org/wiki/Hilbert\_transform

\bibitem{Kang}H.~C. Kang, J.~Maser, G.~B. Stephenson, C.~Liu,
R.~Conley, A.~T. Macrander, and S.~Vogt, ``Nanometer linear focusing
of hard x-rays by a multilayer Laue lens,'' \emph{Phys. Rev. Lett.}
\textbf{96}, 127401 (2006). doi.org/10.1103/PhysRevLett.96.127401

\bibitem{Conley}H.~C. Kang, G.~B. Stephenson, C.~Liu, R.~Conley,
R.~Khachatryan, M.~Wieczorek, A.~T. Macrander, H.~Yan, J.~Maser,
J.~Hiller, and R.~Koritala, ``Sectioning of multilayers to make
a multilayer Laue lens,'' \emph{Rev. Sci. Instrum.} \textbf{78},
046103 (2007). doi.org/10.1063/1.2713439

\bibitem{Yan}H.~Yan, R.~Conley, N.~Bouet, and Y.~S. Chu, ``Hard
x-ray nanofocusing by multilayer Laue lenses,'' \emph{J. Phys. D:
Appl. Phys.} \textbf{47}, 263001 (2014). doi.org/10.1088/0022-3727/47/26/263001

\bibitem{Mass_attenuation_coefficient_table}J. H. Hubbell and S.
M. Seltzer, \textquotedbl Tables of X-Ray Mass Attenuation Coefficients
and Mass Energy-Absorption Coefficients,\textquotedbl{} National Institute
of Standards and Technology (1995), https://physics.nist.gov/PhysRefData/XrayMassCoef/ElemTab/z79.html.

\bibitem{Krajcsi}V. Krajcsi, P. Karvinen, S. S. Parvanabaideh, A.
Diaz, and C. David, \textquotedbl Efficiency of high-aspect-ratio
multilayer Laue lenses for the soft-to-hard X-ray transition,\textquotedbl{}
Opt. Express 24, 18428 (2016).

\bibitem{Medjanik}K. Medjanik, S. V. Babenkov, S. Chernov, D. Vasilyev,
B. Sch{\"{o}}nhense, C. Schlueter, A. Gloskovskii, Yu. Matveyev, W. Drube,
H. J. Elmers, and G. Sch{\"{o}}nhense, \textquotedbl Progress in HAXPES
performance combining full-field k-imaging with time-of-flight recording,\textquotedbl{}
J. Synchrotron Rad. 26, 1996\textendash 2012 (2019)

\bibitem{Henke}B. L. Henke, E. M. Gullikson, and J. C. Davis, \textquotedbl X-ray
interactions: photoabsorption, scattering, transmission, and reflection
at E=50-30000 eV, Z=1-92,\textquotedbl{} Atomic Data and Nuclear Data
Tables, vol. 54, no. 2, pp. 181-342 (1993).

\bibitem{3DRadon}Amir Averbuch and Yoel Shkolnisky, Appl. Comput.
Harmon. Anal. 15 (2003) 33\textendash 69

\bibitem{resolution}Goodman, J. W., Introduction to Fourier Optics,
4th ed., Roberts \& Company Publishers, 2017.

\bibitem{reconstruction}Algebraic Reconstruction Algorithms, Purdue
University, https://engineering.purdue.edu/\textasciitilde malcolm/pct/CTI\_Ch07.pdf

\bibitem{weak_WIKI}https://en.wikipedia.org/wiki/Weak\_measurement

\bibitem{ahanorov}Yakir Aharonov, David Z. Albert, and Lev Vaidman,
VOLUME 60, NUMBER 14 PHYSICAL REVIEW LETTERS How the Result of a Measurement
of a Component of the Spin of a Spin- 2 Particle Can Turn Out to be
100.

\bibitem{ludeen} Jeff S. Lundeen, Brandon Sutherland, Aabid Patel,
Corey Stewart \& Charles Bamber,''Direct measurement of the quantum
wavefunction'', 1 8 8 \textbar{} N A T U R E \textbar{} V O L 4 7
4 \textbar{} 9 J U N E 2 0 1 1 https://www.nature.com/articles/nature10120

\bibitem{kocsis}Sacha Kocsis, Boris Braverman, Sylvain Ravets, Martin
J. Stevens, Richard P. Mirin, L. Krister Shalm, Aephraim M. Steinberg\dag ,
``Observing the Average Trajectories of Single Photons in a Two-Slit
Interferometer'', Science . 2011 Jun 3;332(6034):1170-3. doi: 10.1126/science.1202218.

\end{thebibliography}
\end{document}